\documentclass[preprint]{aastex}  
\usepackage{graphicx,epsfig,times}

\shortauthors{McIntosh et al.}
\shorttitle{Ongoing Assembly of Massive Galaxies}

\newcommand{\sersic}{{S\'{e}rsic }}

\newcommand{\kms}{km s$^{-1}$}
\newcommand{\mpsa}{mag arcsec$^{-2}$}

\slugcomment{{\sc Draft: } \today }

\begin{document}



\title{Ongoing Assembly of Massive Galaxies by Major Merging in Large Groups and Clusters from the SDSS}

\author{Daniel H.\ McIntosh$^{1,2}$, Yicheng Guo$^1$, Jen Hertzberg$^{1,3}$, Neal Katz$^1$, H. J. Mo$^1$}
\author{Frank C. van den Bosch$^4$, Xiaohu Yang$^{5,6}$}
\affil{$^1$ Astronomy Department, University of Massachusetts, 
710 N. Pleasant St., Amherst, MA 01003, USA}
\affil{$^2$ email:
{\texttt dmac@hamerkop.astro.umass.edu}}
\affil{$^3$ Current address: Stony Brook University, Marine Sciences Research Center, Stony Brook, NY 11794-5000}
\affil{$^4$ Max-Planck-Institut f\"ur Astronomie,
K\"onigstuhl 17, D-69117 Heidelberg, Germany}
\affil{$^5$ Shanghai Astronomical Observatory, the Partner Group of MPA, 
Nandan Road 80, Shanghai 200030, China}
\affil{$^6$ Joint Institute for Galaxy and Cosmology of Shanghai Astronomical
Observatory and University of Science and Technology of China}

\begin{abstract}
We investigate the incidence of major mergers creating
${\rm M}_{\rm star}>10^{11} {\rm M}_{\sun}$ galaxies in the dense environments
of present-day groups and clusters more massive than 
${\rm M}_{\rm halo}=2.5\times10^{13}{\rm M}_{\sun}$. We identify 38 pairs of
massive galaxies with mutual tidal interaction signatures
selected from $>5000$ galaxies with
${\rm M}_{\rm star}\geq5\times10^{10} {\rm M}_{\sun}$ that reside in
a halo mass-limited sample of 845 groups. 
We fit the images of each galaxy pair
as the line-of-sight projection of symmetric models and identify mergers
by the presence of residual asymmetric structure associated with both
progenitors, such as nonconcentric isophotes, broad and diffuse tidal
tails, and dynamical friction wakes.
At the resolution and sensitivity of the SDSS, such mergers are found
in 16\% of the high-mass, galaxy-galaxy pairs with $\leq1.5$ $r$-band magnitude
differences and $\leq30$ kpc projected separations. Relying on automated
searches of major pairs from the SDSS spectroscopic galaxy sample will
result in missing 70\% of these mergers owing to spectroscopic
incompleteness in high-density regions.
We find that 90\% of
these mergers are between two nearly equal-mass progenitors with red-sequence
colors and centrally-concentrated morphologies, in agreement with
numerical simulations that predict that an important mechanism for the
formation of massive elliptical galaxies is the dissipationless (gas-poor or
so-called dry) major merging of spheroid-dominated galaxies.
We identify seven additional ${\rm M}_{\rm star}>10^{11} {\rm M}_{\sun}$
mergers with disturbed morphologies and semi-resolved double nuclei.
Mergers at the centers of massive groups are more common than between
two satellites, but both types are morphologically indistinguishable
and we tentatively conclude that the latter are likely located at the dynamical
centers of large subhalos that have recently been accreted by their 
host halo, rather than the centers of distinct halos seen in projection.
We find that the frequency of central and satellite merging 
diminishes with group mass in a manner that is consistent with dynamical
friction.
Based on reasonable assumptions, the centers of these massive halos
are gaining stellar mass at a rate of 1--9\% per Gyr on average.
Compared to the merger rate for the overall population of luminous red
galaxies, we find that the rate is 2--9 times greater when restricted
to these dense environments.
Our results imply that the massive end of the galaxy population
continues to evolve hierarchically at a measurable level, and that
the centers of massive groups are the preferred
environment for the merger-driven assembly of massive ellipticals.
\end{abstract}

\keywords{galaxies: evolution --- galaxies: fundamental parameters (luminosities, stellar masses, radii) --- galaxies: general --- surveys}


\section{Introduction}

Understanding the formation of the most-massive galaxies
(${\rm M}_{\rm star}>10^{11}{\rm M}_{\sun}$)
remains an important challenge in astrophysics.
The tip of the stellar mass function is dominated by elliptical
galaxies with intrinsically spheroidal mass distributions that
are supported by anisotropic stellar motions \citep{kormendy96,burstein97}.
Numerical simulations have long demonstrated that
``major'' mergers between smaller galaxies of comparable mass could
produce the observed
shapes and dynamics of ellipticals 
\citep{toomre77,barnes96,naab03,cox06}. Moreover, massive ellipticals are found
in greater abundance in high-density structures like large groups and 
clusters of galaxies
\citep[e.g.,][]{dressler80a,postman84,hashimoto99,smith05},
which naturally grow through the hierarchical merging of dark-matter halos
over cosmic time as expected in the $\Lambda$CDM cosmological model 
\citep{blumenthal84,davis85,cole00}.
There is, therefore, a clear expectation for galaxy-galaxy and halo-halo
merging to be physically linked \citep{maller06,hopkins06b,delucia07}. Indeed,
modern galaxy formation models predict that massive ellipticals form by
major dissipationless (so-called ``dry'') merging of likewise 
spheroidal and gas-poor progenitors \citep{boylan06,naab06a},
that a large fraction of today's massive ellipticals had their last
major merger since redshift $z=0.5$ \citep[e.g.,][]{delucia06},
and that the most-massive systems form
at the centers of large dark-matter halos 
\citep{dubinski98,aragon98}.
Yet, direct evidence for the major-merger 
assembly of massive galaxies at present times has 
been lacking, and finding such systems is needed to place 
constraints on their rates, progenitor properties,
and environmental dependencies.
To this end we look for close pairs of massive interacting galaxies
within a complete and well-defined sample
of over 5000 galaxies with $z\leq0.12$ and
${\rm M}_{\rm star}\geq5\times10^{10}{\rm M}_{\sun}$,
selected from galaxy groups in the Sloan Digital Sky Survey (SDSS) with
dark-matter halo masses above
${\rm M}_{\rm halo}=2.5\times10^{13}{\rm M}_{\sun}$.


Ellipticals galaxies make up the bulk of the massive end of the
red-sequence population with optical colors indicative of their
non-star-forming and old stellar nature.
Despite a quiet star-formation history
over the last 6--8 billion years \citep{bell05a}, the total stellar mass
density on the red sequence has roughly doubled over this
interval \citep{bell04b,blanton06,borch06,faber07,brown07} 
and now accounts for more than half of
the present-day budget \citep{hogg02,bell03b}, 
providing strong observational evidence
for the ongoing hierarchical growth
of the massive galaxy population. These results were derived from
red galaxy number densities over a wide range of
stellar masses above and below 
$10^{11}{\rm M}_{\sun}$.
Owing to the scarcity of the highest-mass galaxies, cosmic variance,
and systematic uncertainties in stellar mass estimates, any increase
in the number density of 
${\rm M}_{\rm star}>10^{11}{\rm M}_{\sun}$ galaxies is poorly
constrained, resulting in
controversy over whether this population
has continued to grow slowly \citep[e.g.,][]{brown07}
or has been effectively static \citep[e.g.,][]{scarlata07}, since $z\sim1$.

Besides number density evolution, mergers of
sufficiently massive galaxies could provide a more clear indication
for some continued stellar mass growth in the high-mass galaxy population.
The existence of a handful of massive red mergers over the redshift
interval $0.1<z<0.9$ \citep{vandokkum99,tran05,bell06a,lotz06,rines07} 
proves that the growth is non-zero
at high stellar masses and implies that this mechanism does contribute
to the assembly of galaxies at the top of the food chain.
Yet, the importance of this process and
the related rate of mass growth are highly uncertain given the tiny
samples over this large cosmic time interval.
Indirect measures such as the presence of faint tidal debris or shells around
many local massive ellipticals \citep{vandokkum05,mihos05}, the
isophotal properties of giant ellipticals \citep{kang07},
the lack of evolution of the stellar mass-size relation of red spheroids
since $z=1$ \citep{mcintosh05b}, and the lack of
morphological evolution on the red sequence since $z=0.7$ \citep{bell04a}
provide a variety of limits to the importance of dissipationless mergers.
Perhaps the most powerful method for obtaining estimates for the
stellar mass growth rate via major merging is
based on small-scale clustering statistics
that provide an accurate measurement of close pair frequencies in real space
\citep{masjedi06,bell06b,masjedi07}. However, this method likely yields an 
overestimate
of the merger frequency because it assumes that all close pairs will merge.
All estimates of merger-driven growth rates are limited by uncertainties
in the time interval over which a pair will merge, and over what duration
an object could be identified as interacting.
\citet{masjedi07} find a very small
growth rate (1--2\% per Gyr) at $z\sim0.25$ for major mergers involving
at least one progenitor drawn from 
the SDSS Luminous Red Galaxy sample \citep[LRG,][]{eisenstein01};
LRGs have typical masses of several times $10^{11}{\rm M}_{\sun}$.
To date there remains no direct evidence of ongoing merger-driven 
assembly of massive galaxies at $z<0.1$, and the LRG result implies
that this formation process is no longer important.
These facts motivate a thorough search for the existence/nonexistence
of ongoing examples in the present-day universe.

While the aforementioned statistical
method for finding close physical pairs is powerful, it 
does not isolate actual merging systems and thus
provides no information on the progenitor properties of massive
merger remnants. Recent numerical simulations and models make a range of predictions
regarding the progenitor morphologies at the time of the last major
merger \citep{khochfar03,naab06a,kang07}, yet robust observational constraints
are missing for ${\rm M}_{\rm star}>10^{11} {\rm M}_{\sun}$ systems.
Many studies have identified major-merger candidates by either
close pairs \citep{carlberg94,patton00,carlberg00,patton02,bundy04,lin04} 
or disturbed morphologies \citep{lefevre00,conselice03,lotz06}, but
these samples mostly contain major mergers between
lower-luminosity galaxies that tend to be gas-rich spiral disks.
Numerical simulations show that such dissipative merging of
disk galaxies will not produce massive pressure-supported ellipticals
\citep[e.g.,][]{naab06d}.
As mentioned above, only circumstantial evidence and a small number 
of red galaxy pairs with $z<0.9$ support the existence of mergers likely
to produce massive ellipticals. Our understanding of the progenitors
is therefore very limited. Here we present a thorough census of
38 massive merger pairs from SDSS, providing an order-of-magnitude
increase in the number of such detections at $z<0.5$ and allowing
an improved understanding of their progenitor properties.

While many estimates of major merger rates are found in the literature,
to date no measure of the environmental dependence of merger-driven
mass growth has been attempted.
In the standard cosmological model, there is a trade off between the
expansion of the universe and the gravitational collapse of dark
and luminous matter.
Therefore, the rate at which stellar mass is assembled at the centers of
the largest dark matter halos over recent cosmic history is a
fundamental aspect of the ongoing formation of large-scale structure,
and the rate that high-mass galaxies form by mergers as a function
of halo mass constrains galaxy formation theories.
Some theories predict that the mergers producing massive
ellipticals occur preferentially in groups rather than in high-density cluster 
or low-density field environments because the smaller velocity dispersions allow
more galaxy interactions \citep{cavaliere92}; also dynamical friction
is more efficient in lower-mass halos \citep[e.g.][]{cooray05d}. Others
predict that the brightest cluster galaxies (BCGs) grow by hierarchical merging
(``galactic cannibalism'') at the
centers of the dark-matter potential wells of large clusters 
\citep{ostriker75,merritt85,dubinski98,cooray05d}. A handful of low-redshift
BCGs show multiple nuclei suggesting cannibalism in the form of multiple
minor mergers \citep{lauer88}, but there are no observations of major mergers 
at the centers of clusters.
In this paper we make use of the
statistically large SDSS group catalog \citep{yang05a,weinmann06a}
to show that major mergers occur in present-day dense environments,
and to explore the halo-mass dependence and central/satellite identity of
merger-driven massive galaxy assembly.

Throughout this paper we calculate comoving distances in the $\Lambda$CDM
concordance cosmology with $\Omega_{\rm m} = 0.3$,
$\Omega_{\Lambda} = 0.7$, and assume a Hubble constant of
$H_0 = 70 $\,km\,s$^{-1}$\,Mpc$^{-1}$. SDSS magnitudes are in the
AB system.

\section{Sample Selection}
\label{sec:data}

We make use of public catalogs derived from the SDSS \citep{york00}
Data Release Two \citep[DR2,][]{abazajian04}, which includes 
spectroscopic and imaging coverage of more than 2600 square degrees.
The $ugriz$ passband imaging \citep{fukugita96,gunn98,gunn06},
precise photometry \citep{hogg01,smith02,ivezic04,tucker06},
image processing \citep{lupton02}, astrometric calibration \citep{pier03},
and spectroscopy \citep{strauss02,blanton03d}
of the SDSS provides a powerful database
for detailed studies of the galaxy population from the local cosmos.
We exploit the large-number statistics of the SDSS to search for
elusive pairs of massive galaxies undergoing major merging in
dense group and cluster environments. As described in detail below,
our sample selection consists of (1)
a complete and mass-limited set of large dark-matter
halos drawn from the SDSS DR2
group catalog\footnote{Public access to the group catalog
is at {\texttt http://www.astro.umass.edu/$\sim$xhyang/Group.html}.}
\citep{yang05a,weinmann06a}, (2) the subset of
massive galaxy pairs within these groups that meet the stellar mass
criteria of ${\rm M}_1+{\rm M}_2\geq10^{11} {\rm M}_{\sun}$,
and (3) the identification of merger candidates among the massive pairs.

\subsection{Massive Halos from the SDSS Group Catalog}

With large surveys of spectroscopic redshifts and imaging data, 
astronomers are for the first time able to study galaxies according to 
their membership and position within dark-matter halos (i.e., galaxy groups).
Using the halo-based group finder of \citet{yang05a}, \citet{weinmann06a}
extracted groups
from an initial sample of 184,425 galaxies with $0.01\leq z\leq0.20$
and better than 70\% redshift completeness drawn from the New York University
Value-Added Galaxy Catalog
\citep[NYU-VAGC,][]{blanton05}.
The NYU-VAGC provides improved processing and
additional parameters for the SDSS spectroscopic Main galaxy sample 
\citep{strauss02},
which has an extinction-corrected $r=17.77$ magnitude limit.

The halo-based  group finder of \citet{yang05a} has  been optimized to
group galaxies  according to  their common dark-matter halo,  and has
been thoroughly  tested using mock galaxy redshift  surveys.  Briefly,
the group finder starts  with a friends-of-friends algorithm to define
potential groups and their  centers. Any isolated, bright galaxies not
assigned  to  a  potential  group  are  added  as  likely  centers  of
additional groups. The total group luminosity is
converted into an  estimate for  the group  mass using  an assumed
mass-to-light (M/L) ratio. From this  mass estimate, the radius and velocity
dispersion of  the corresponding dark-matter halo  are estimated using
the virial equations,  which in turn are used  to select group members
in redshift  space.  This method  is iterated until  group memberships
converge.  In \citet{yang05a}, the performance of this group finder has
been tested in terms of completeness of true members and contamination
by  interlopers, using  detailed  mock galaxy  redshift surveys.   The
average completeness  of individual groups  was found to be  $\sim 90$
percent, with  only $\sim  20$ percent interlopers.   Furthermore, the
resulting  group catalogue  is insensitive  to the  initial assumption
regarding the M/L ratios.

As described in \citet{weinmann06a},
halo masses  for each identified  group were estimated using  the total
group luminosity, $L_{\rm group}$, defined as the summed luminosity of
all group members.   The motivation behind this is  that one naturally
expects the group luminosity to be strongly correlated with halo mass.
Because of the  flux limit of the SDSS,  two identical groups observed
at different  redshifts will have  a different $L_{\rm  group}$.  This
bias was circumvented by using  $L_{19.5}$ instead, which is defined as
the luminosity of all group members brighter than $^{0.1}M_r = -19.5 +
5\log  h$.  The  relation between  $L_{19.5}$ and  $L_{\rm  group}$ was
calibrated using groups  with $z \leq 0.09$, which  corresponds to the
redshift for which a galaxy with  $^{0.1}M_r = -19.5 + 5\log h$ has
$r=17.77$, the magnitude limit of the survey.
For groups  with $z >  0.09$, this `local' calibration  between $L_{\rm
group}$ and $L_{19.5}$ was used to estimate the latter.
Finally, under the assumption  that there  is  a one-to-one relation 
between $L_{19.5}$  and halo mass, and using the halo 
mass  function corresponding  to  a flat  $\Lambda$CDM cosmology  with
$\Omega_m=0.3$  and $\sigma_8=0.9$, the  halo mass  of a  given group,
${\rm M}_{\rm halo}$, then follows from matching the number density of 
groups brighter
(in terms  of $L_{19.5}$) than the group in consideration  to that of
halos  more  massive  than  ${\rm M}_{\rm halo}$.  
Detailed tests  with  mock  galaxy
redshift surveys have  shown that this method results  in group masses
that are more reliable than  those based on the velocity dispersion of
the  group members,  especially when  the number  of group  members is
small.

In \citet{weinmann06a},  this group finder was applied  to the NYU-VAGC
associated with  the SDSS  DR2, which yielded  halo masses  for 53,229
groups     spanning     $11.8<\log_{10}({\rm    M}_{\rm     halo}/{\rm
  M}_{\sun})<15.5$ and containing 92,315  galaxies. As a result of the
method used to assign the  group masses, the completeness of the group
catalog  depends on  both  halo  mass and  redshift.   In detail,  the
catalog is complete for groups with $\log_{10}({\rm M}_{\rm halo}/{\rm
  M}_{\sun})>11.8622$ to  $z=0.06$, $\log_{10}({\rm M}_{\rm halo}/{\rm
  M}_{\sun})>12.1933$   to  $z=0.12$,   and   $\log_{10}({\rm  M}_{\rm
  halo}/{\rm M}_{\sun})>13.0877$ to $z=0.20$.

For  our   analysis,  this   group  catalog  provides   two  important
environmental measures for every member galaxy: (1) an estimate of the
virial mass (${\rm M}_{\rm halo}$) of the dark-matter halo in which the
galaxy  resides,  and (2)  a  distinction  between  central (CEN)  and
satellite (SAT)  galaxies. Throughout, a CEN galaxy  is defined as
the  brightest  member  of  its  group.  As  discussed  in  detail  in
\citet{weinmann06a}, these quantities allow more physically-meaningful
discussions of  the dependencies  of galaxy properties  on environment
than do projected number densities.

\begin{figure*}
\center{\includegraphics[scale=0.8, angle=0]{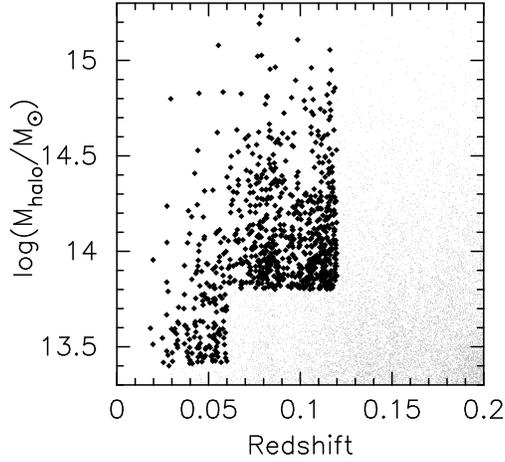}}
\caption[]{Distribution of halo mass and redshift for massive groups
identified by \citet{weinmann06a} in the SDSS DR2.
Small grey points show all 12,552 groups with 
${\rm M}_{\rm halo}>2\times10^{13}{\rm M}_{\sun}$; for $z\leq0.12$
there are 2666 groups above this halo mass cut, the majority of which
contain only 1 or 2 spectroscopic members. Black diamonds show our 
halo-mass and volume-limited selection of 845 groups (see text for details)
that we use to search for major pairs of massive galaxies.
There are 176 groups with $0.01<z\leq0.06$ (vol. I) and 
${\rm M}_{\rm halo}>2.5\times10^{13}{\rm M}_{\sun}$, and 669 with
$0.06<z\leq0.12$ (vol. II) 
and ${\rm M}_{\rm halo}>6.3\times10^{13}{\rm M}_{\sun}$.
\label{fig:sub_selection}}
\vspace{-0.2cm}
\end{figure*}

We combine two volume-limited samples
defined by the halo construction and
completeness described above: (I) $0.01<z\leq0.06$ and (II) $0.06<z\leq0.12$.
We exclude halos with $0.12<z\leq0.20$ to avoid resolution limitations.
At $z=0.12$, the SDSS resolution of $1.4\arcsec$ corresponds to 3 kpc, thus
fairly massive galaxies will be only semi-resolved. Moreover, 
unreliable photometry is known to occur in SDSS for galaxies separated by
$<3\arcsec$ \citep{masjedi06}, which corresponds to 7--10 kpc over the
$0.12<z\leq0.20$ interval. We find many close pairs with physical separations
less than 10 kpc, thus our redshift cut avoids selecting
a large fraction of close pairs with poor photometry.
Within the two redshift slices we further limit our selection to halos
that have at least three spectroscopic members to allow for
a complete search of massive pairs associated with either CEN or
SAT galaxies.
This restricts our final sample to all SDSS DR2 groups with masses of
$\log_{10}({\rm M}_{\rm halo}/{\rm M}_{\sun})\geq13.4$ in volume I, and
$\log_{10}({\rm M}_{\rm halo}/{\rm M}_{\sun})\geq13.8$ in II.
Hence, our selection is halo mass-limited at values significantly larger
than the group catalog completeness limits.
We plot the halo mass and redshift distribution
of our final sample in Figure \ref{fig:sub_selection}, which
contains 845 groups with masses ranging from one-tenth
to ten times that of the Virgo cluster.

\subsection{Massive Galaxy Pairs}
\label{sec:mgp}

The primary goal of our study is to find whether evidence exists 
for the major-merger assembly of massive 
(${\rm M}_{\rm star}>10^{11} {\rm M}_{\sun}$) galaxies in dense environments.
We approach this by first inspecting all massive galaxies belonging to
the halo mass-limited selection of
845 groups to identify those systems that have a major companion
(mass ratio between 4:1 and 1:1)
within a projected separation of 30 kpc. It is important to keep in mind
that these companions are neither restricted to be in the SDSS spectroscopic
Main sample nor in the group catalog.
We then use an image decomposition technique, 
as described in \S \ref{sec:inter},
to identify the pairs that exhibit signs of tidal
interaction associated with an ongoing merger.

We estimate stellar masses (${\rm M}_{\rm star}$) for all group members
using the \citet{bell03b} stellar M/L ratios as follows:
\begin{equation}
\log_{10}({\rm M}_{\rm star}/{\rm M}_{\sun})=-0.306 + 1.097^{0.0}(g-r) - 0.15 - 0.4(^{0.0}M_r-4.67) ,
\label{eq:1}
\end{equation}
where the constant 0.15 corrects to a \citet{kroupa01} IMF, and
$^{0.0}g$ and $^{0.0}r$ are Petrosian magnitudes from the NYU-VAGC
(random uncertainties $<0.03$ mag)
shifted to the $z=0$ rest-frame using \citet{blanton03b} $K$-corrections and
corrected for Milky Way extinction using the \citet{schlegel98} dust maps.
We subtract $0.1$ magnitude to correct for the flux known to be missing
from galaxies with early-type morphologies \citep{blanton03c}.
We use the $r$-band central-light concentration ($R_{90}/R_{50}$),
defined by the ratio of the radii containing 90\% and 50\% of the Petrosian
flux, to coarsely 
separate early-type ($R_{90}/R_{50}\geq2.6$; spheroid-dominated) and 
late-type ($R_{90}/R_{50}<2.6$; disk-dominated) galaxies as others
have with SDSS data
\citep[e.g.,][]{strateva01,hogg02,bell03b,kauffmann03b}.
The \citet{bell03b} color-based stellar M/L ratios have 20\% random uncertainties
and a 0.10-0.15 dex systematic error caused by a combination of effects
including dust, stellar population ages,
and bursts of star formation.
The characteristic stellar mass
of the local galaxy mass function from \citet{bell03b}
is ${\rm M}^{\ast}=7.24\times10^{10} {\rm M}_{\sun}$.

\begin{figure*}
\center{\includegraphics[scale=0.8, angle=0]{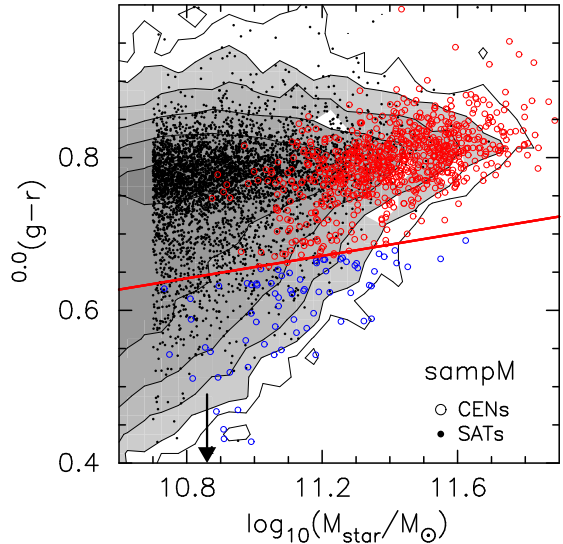}}
\caption[]{
Rest-frame optical color versus stellar mass
plot for our selection of massive members (sampM) from
a halo mass-limited sample of large SDSS DR2 groups.
Grayscale contours show all SDSS (DR2) galaxies with 
$0.01<z\leq0.12$;
each contour represents a 3-fold increase in the number of galaxies. The
solid red line is the red/blue sequence separation we adopt from
\citet{weinmann06a}.
Blue and red circles denote the subset of 845 CEN galaxies,
and black solid points denote the 4531 satellites.
The vertical arrow indicates ${\rm M}^{\ast}$ from \citet{bell03b}.
\label{fig:cm.sampM}}
\vspace{-0.2cm}
\end{figure*}

To find major mergers between two galaxies with mass ratios $\leq4:1$,
we start with all 5376 group members more massive than
${\rm M}_{\rm star}=5\times10^{10} {\rm M}_{\sun}$ 
(hereafter {\bf sampM}) and note that
this mass limit is the minimum for which an equal-mass merger will
produce a $10^{11} {\rm M}_{\sun}$ remnant. 
We plot the color versus stellar mass distribution of sampM
in Figure \ref{fig:cm.sampM}. The contours represent all SDSS
DR2 Main galaxies with $z\leq0.12$.
The halo-mass-limited sample of 845 CEN galaxies from sampM
are shown as red and blue circles separated by the red/blue sequence boundary
from \citet{weinmann06a}, modified to $z=0$ and
$H_0 = 70 $\,km\,s$^{-1}$\,Mpc$^{-1}$. The 4531 SATs from sampM are plotted
as black solid points.
Not surprising, the vast majority of massive
galaxies in high-mass groups (both CEN and SAT) have red-sequence colors.
We compare the massive galaxy content of sampM with that of the $z\leq0.12$
DR2 volume in Table \ref{mstar_counts}.

\begin{figure*}
\center{\includegraphics[scale=0.7, angle=0]{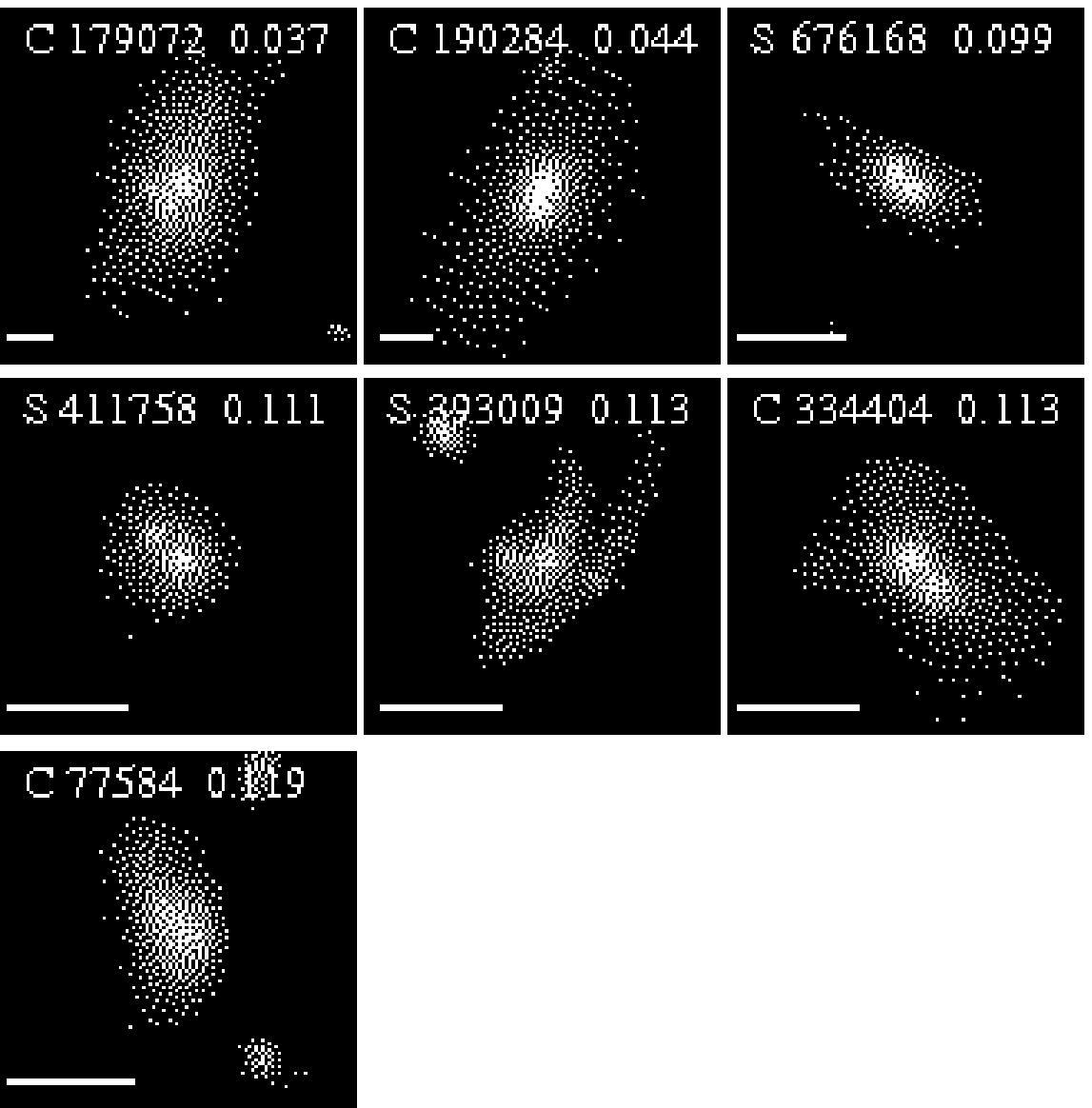}}
\caption[]{Seven additional mergers identified only by strong morphological
disturbances, not by close major companions. Images are $60\times60$ kpc
cutouts of $gri$ combined color images with fixed sensitivity scaling
downloaded from the SDSS Image List Tool. We distinguish CEN (C) from SAT (S)
mergers, and we include the NYU ID and the redshift,
at the top of each panel. The horizontal white line shows $10\arcsec$ in each
panel.
\label{fig:mergers}}
\vspace{-0.2cm}
\end{figure*}

We use the SDSS Image List
Tool\footnote{Available from the SDSS SkyServer Tools at {\texttt http://cas.sdss.org/astro/en/tools/chart/list.asp}.}
as a virtual observatory to visually examine an $80\times80$ kpc region
centered on each massive galaxy in sampM, which allows
us to view the entire extent of both galaxies in a 30 kpc pair.
Although more time-consuming, this method ensures that we find all major
companions including those without SDSS spectroscopic data.
In addition, our examination allows the identification of individual
(non-pair) sources with highly disturbed morphologies suggestive of
ongoing major mergers, which cannot be found with automated pair selection.
We find seven morphologically-identified mergers that have semi-resolved double
nuclei with projected separations too close to be accurately deblended
by the SDSS (Fig. \ref{fig:mergers}).

We find that 221 massive galaxies in sampM have a major companion
with a projected separation of $d_{12}\leq30$ kpc (centroid-to-centroid).
Operationally, we use an apparent 
$r$-band magnitude difference of $\vert \Delta r_{12}\vert \leq1.5$,
corresponding to mass ratios $\leq4:1$ assuming a constant M/L ratio,
to identify major companions both with and without spectroscopic data.
Throughout this paper we use the following
designations for projected pairs: galaxy number 1 is from sampM and
galaxy number 2 is its projected companion, regardless of relative brightness
or mass. In the cases where both galaxies have spectroscopic redshifts and are
massive enough to be included in sampM, galaxy 1 is the primary
(i.e., brightest) member and we remove from further analysis 
the duplicated pair initiated on galaxy number 2.
The SDSS spectroscopy is known to be about 8\% incomplete overall, 
independent of galaxy luminosity. The main source of incompleteness
results from the $55\arcsec$ minimum separation for fiber placement
(i.e., ``fiber collisions'')
in the mechanical spectrograph \citep{blanton03d}. This selection effect 
leads to a slight
systematic under representation in regions of high galaxy number density
\citep{hogg04}, such as in massive groups and clusters.
Less than one third of the 221 pairs have spectra for both galaxies, and
thus, redshifts for galaxies number 1 and 2 (i.e., spec-spec pairs).
In what follows we will show that an important fraction of all
pair-identified massive mergers have only one spectroscopic progenitor
(i.e., spec-phot pairs).

Close pairs of galaxies are used often to infer information about galaxy merging
\citep[e.g.][]{patton00,carlberg00,lefevre00,patton02,lin04,bundy04,bell06b}.
These studies use a range of definitions, which usually include tight limits on
both velocity and projected spatial separations (typically $<500$ \kms\
and 10--50 kpc), and do not use further knowledge such
as the halo mass or the position relative to the group center.
Our choice of $d_{12}\leq30$ kpc separations
is rather arbitrary and we have no way of knowing ab initio whether or not
it will include all massive pairs that show obvious signs of interaction.
Owing to our larger $80\times80$ kpc field of view, we effectively search 
within a projected radius of 40 kpc around each galaxy in sampM, which
enable us to find three additional wide-separation ($d_{12}>30$ kpc)
pairs that exhibit strong merging signatures.
The maximum projected separation of the additional mergers is 37 kpc.
The minute frequency of $30<d_{12}<40$ kpc pairs with strong
tidal signatures in these groups suggests that wider-separation systems
will not be apparent in SDSS-depth imaging data.
We include only pairs with $d_{12}\leq30$ kpc for our projected pair
statistics, but include the three additional merger pairs
in our progenitor and mass assembly statistics.

Our sample includes three pairs with 2--3$\arcsec$ separations, which
are a potential source of systematic bias in our major pair selection.
As mentioned above, 
\citet{masjedi06} showed that the SDSS photometric pipeline boosts
the recovered flux of individual galaxies in very close pairs. For equal-luminosity 
galaxies separated by 5--$20\arcsec$ the excess is only about 5\%, but this
quickly rises to 20\% at $3\arcsec$ separation. Moreover, the pipeline has
trouble deblending very close pairs as is evident in Figure \ref{fig:mergers}.
We do not attempt to separate the progenitors
of these mergers, instead we assume that they represent major mergers
and explicitly state where we include them in our analysis. 


Finally, the subset of 64 major spectroscopic pairs in sampM
allow us the unique opportunity to test
the frequency of interlopers in massive groups. We find that 25\%
of the spectroscopic pairs (2 CEN-SAT, 14 SAT-SAT) are comprised of
projected galaxies in two separate groups with average absolute velocity
separation $\langle \vert \Delta v \vert \rangle=7550$ \kms.
If we limit our analysis to spectroscopic pairs with $d_{12}\leq30$ kpc
and $\langle \vert \Delta v \vert \rangle \leq500$ \kms, 
we find 5\% contamination from interlopers in basic
agreement with \citet{berrier06},
who used mock galaxy catalogs from cosmological simulations to demonstrate
that 10--50 kpc (14--71 kpc in our assumed cosmology) pairs
with less than 500 \kms\ separation
reside in the same dark-matter halo with a low ($5-20\%$) contamination
from projected interlopers.
Overall, the spectroscopic pairs from sampM that live in the same
group have absolute velocity differences spanning 10 to 1560 \kms\ with
means of 260 \kms\ (CEN-SAT) and 360 \kms\ (SAT-SAT). 
Many of these pairs are likely doomed to merge,
yet some may still be chance projections on opposite sides of the
same group.
We feel that the most conservative approach to locating physically interacting
pairs is to look for morphological signs of disturbance, an approach that
we adopt and discuss in the next section.

\subsection{Identifying Major Mergers}
\label{sec:inter}
Besides pair statistics, major galaxy mergers are routinely identified 
by their highly-disturbed appearance 
\citep[e.g.,][]{lefevre00,conselice03,lotz06}. Tidal tails and debris,
multiple nuclei, strong asymmetries, and other morphological peculiarities 
are common features in both observations and in simulations of galaxy collisions
\citep{toomre72,barnes88,barnes92b,dubinski96,barnes96,mihos01}. 
Yet, distinguishing major mergers from lower mass ratio ``minor'' interactions
using morphology alone is fraught with uncertainties. For example, depending
on the orbital geometry, a 10:1 gas-rich merger can result in a more disturbed
morphology than an encounter between two massive ellipticals, which have
broad low-surface brightness features \citep{bell06a}.
We circumvent this issue by
selecting major pairs of massive galaxies first, and then fitting symmetric
models to the light profiles of each galaxy in each major pair
and identifying interaction signatures in the residual (data$-$model) image.
Our methodology is similar in spirit to that of \citet{lauer86b,lauer88},
who modelled BCGs with multiple nuclei as the line-of-sight superpositions of
normal elliptical galaxies.

For each major pair in sampM we use GALFIT \citep{peng02}
to fit the surface photometry of both galaxies and any other
close companions in the SDSS $r$-band image data. For each fit we
use the global background estimate provided in the SDSS image header.
The details of our fitting pipeline developed
for SDSS imaging will be presented in Guo et al. (in prep.).
Asymmetries commonly associated with galaxy mergers
(e.g., tails, bridges, plumes, nonconcentric isophotes, 
diffuse excess structure, and dynamical
friction wakes) are {\it not well fit} by symmetric models
centered on the galaxy. 
Therefore, to isolate and highlight asymmetries in the residual image we
use either a single-component \sersic or a two-component
\sersic bulge plus exponential disk model for each source, depending on whether
or not disk features such as spiral arms, rings, or bars are apparent.
We classify any major pair as a merger if there are
asymmetric residuals brighter than 24.5 \mpsa\
associated with {\it both} progenitor galaxies. All other pairs
are deemed non-interacting. This surface brightness limit was used
in the selection of SDSS spectroscopic target galaxies \citep{strauss02},
and we find that residual features this bright are unambiguous.
 
\begin{figure*}
\center{\includegraphics[scale=0.5, angle=0]{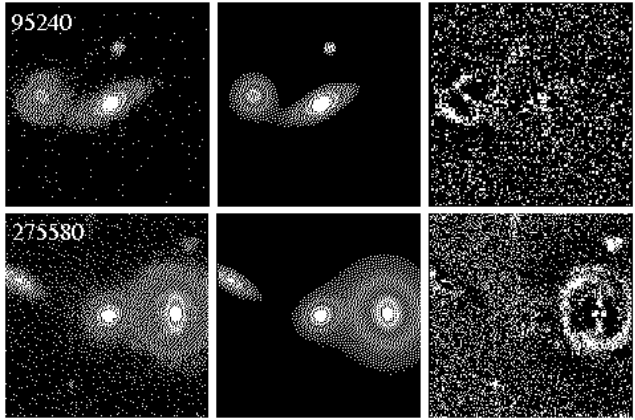}}
\caption[]{
Examples of two galaxy pairs with small projected separations but large
physical separations; i.e., the two galaxies reside in different groups
and are thus not physically associated. {\it Panels:} 
(left) $r$-band SDSS image in arbitrary
false-color, logarithmic scale to highlight low-surface brightness features,
(middle) GALFIT symmetric model profile, and (right) data$-$model residual.
We identify merging galaxies by the presence of asymmetric residual flux
associated with each individual galaxy (see text for details).
These two examples are among the subset of 16 null (interloper) cases,
none of which meet our merger identification criteria. Some interloper
pairs have one galaxy with 
detectable residuals for a variety of reasons other than an
interaction between the two galaxies.
We show the strongest residual cases here to illustrate the most-common cause,
which is spiral structure.
Each image is $80\times80$ kpc and we provide
the NYU-VAGC DR2 identification number (NYU ID) in the upper left.
\label{fig:trueproj1}}
\vspace{-0.2cm}
\end{figure*}

\begin{figure*}
\center{\includegraphics[scale=0.5, angle=0]{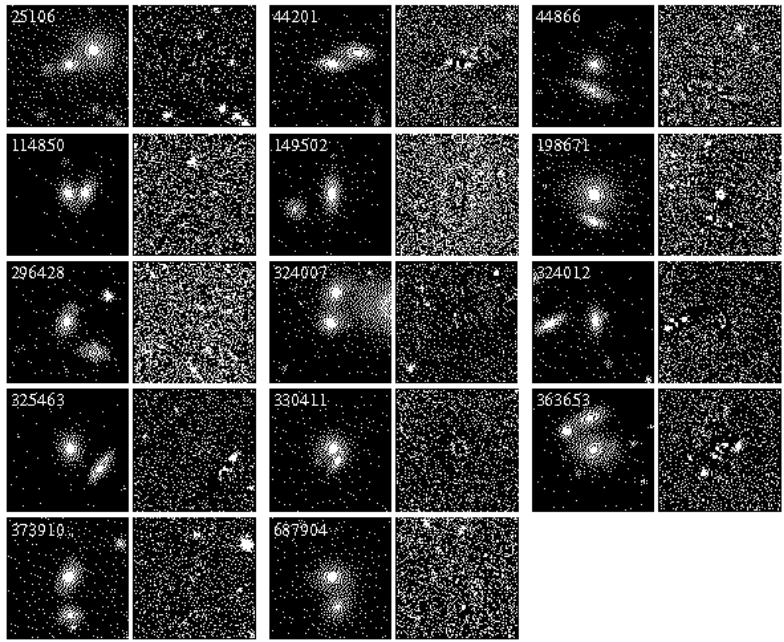}}
\caption[]{
The remaining 14 of 16 major pairs that are interlopers
(individual galaxies in separate groups). No pair exhibits
asymmetric residuals for both galaxies.
The images ($r$-band data and residual with log-scale stretch)
are $80\times80$ kpc and are labeled as
in Figure \ref{fig:trueproj1}.
\label{fig:trueproj2}}
\vspace{-0.2cm}
\end{figure*}

For isolated galaxies not undergoing a major interaction, there are a
number of other explanations for the presence of asymmetric residual flux,
such as lopsided spiral features caused by minor interactions or
large star-forming regions.
The subset of 16 interlopers (\S \ref{sec:mgp}) provides a null sample 
to demonstrate that no major pairs that are simple line-of-sight projections
meet our dual disturbance criteria for identifying mergers.
In Figure \ref{fig:trueproj1}, we show our fitting analysis for
two interlopers that have the strongest detectable asymmetries from spiral
structure (NYU IDs 95240 and 275580). In each case the asymmetry
is associated with only one galaxy of each pair.
None of the remaining 14 interloper pairs in Figure \ref{fig:trueproj2}
exhibit the strong, dual asymmetries that we observe in the merger pairs,
which we describe next.

\begin{figure*}
\center{\includegraphics[scale=0.5, angle=0]{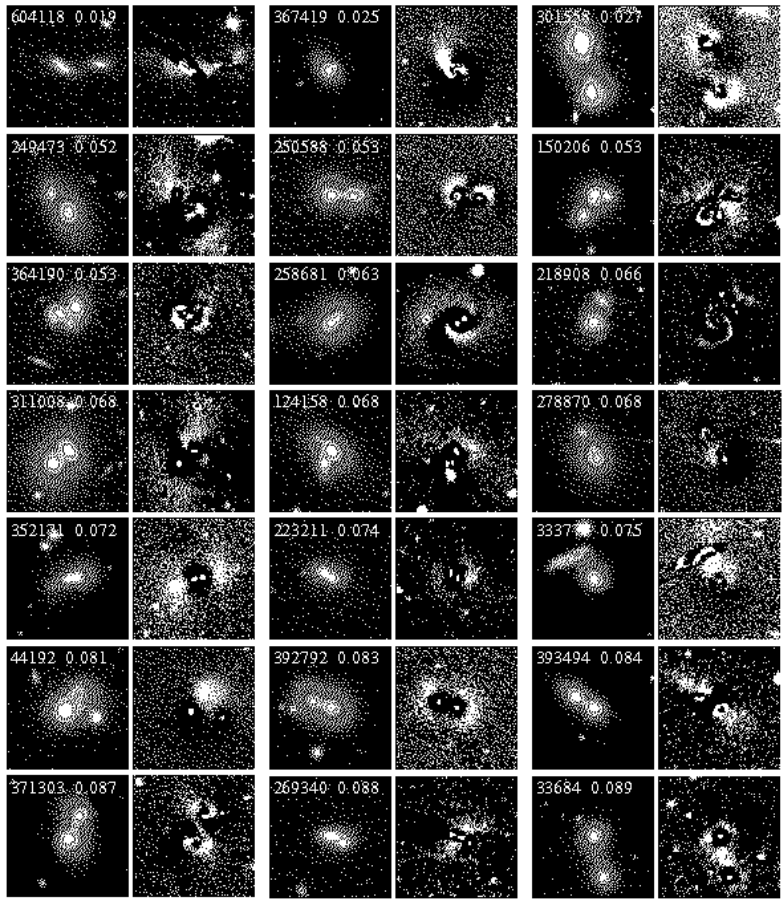}}
\caption[]{See caption for Figure \ref{fig:dpairs2}.
\label{fig:dpairs1}}
\end{figure*}

\begin{figure*}
\center{\includegraphics[scale=0.5, angle=0]{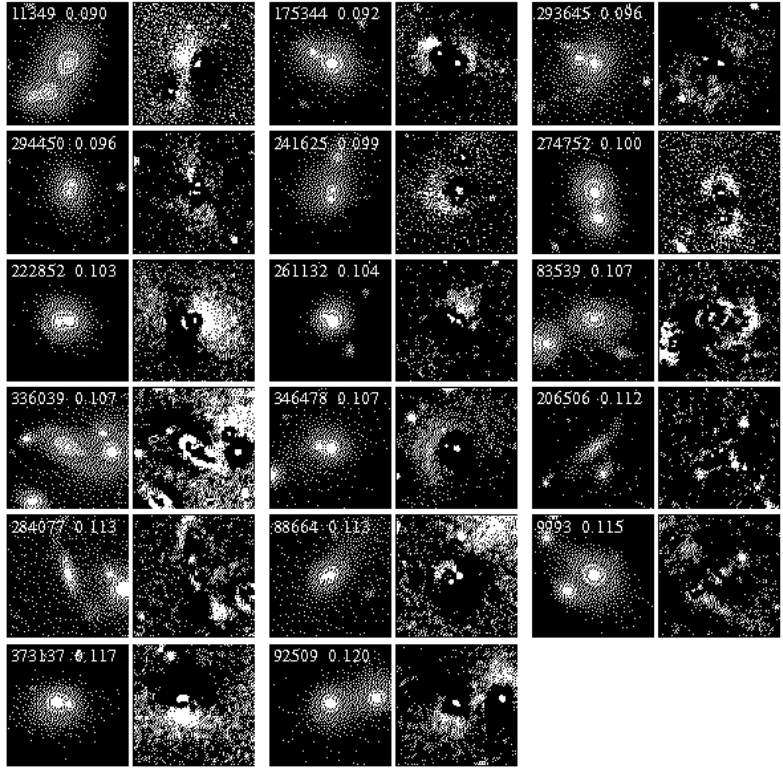}}
\caption[]{
The full sample of 38 major-merger pairs of massive galaxies identified in a
halo mass-limited subset of SDSS DR2 groups with $z\leq0.12$ (sampM).
Three pairs (301558, 83539, and 284077) have projected separations
between $30<d_{12}<37$ kpc.
We identify these merging systems when both
galaxies have asymmetric residual features in excess of 24.5 \mpsa.
Such asymmetries are associated
with tidal signatures (e.g., tails, bridges, plumes, nonconcentric isophotes,
diffuse excess structure, and dynamical friction wakes) of mutual
encounters between two galaxies.  For each pair we provide the $r$-band data
in false color (arbitrary scaling) at the left, and the data$-$model residual
at the right. To highlight low-surface brightness features we
Gaussian smoothed (using a 1 pixel sigma) the residual images of each,
except for 301558, 250588, 364190, 278870, 352171, 333778, 44192, 392792,
371303, 11349, 241625, and 274752. 
All images are $80\times80$ kpc with the NYU ID and 
spectroscopic redshift given.
\label{fig:dpairs2}}
\vspace{-0.2cm}
\end{figure*}

\begin{figure*}
\center{\includegraphics[scale=0.5, angle=0]{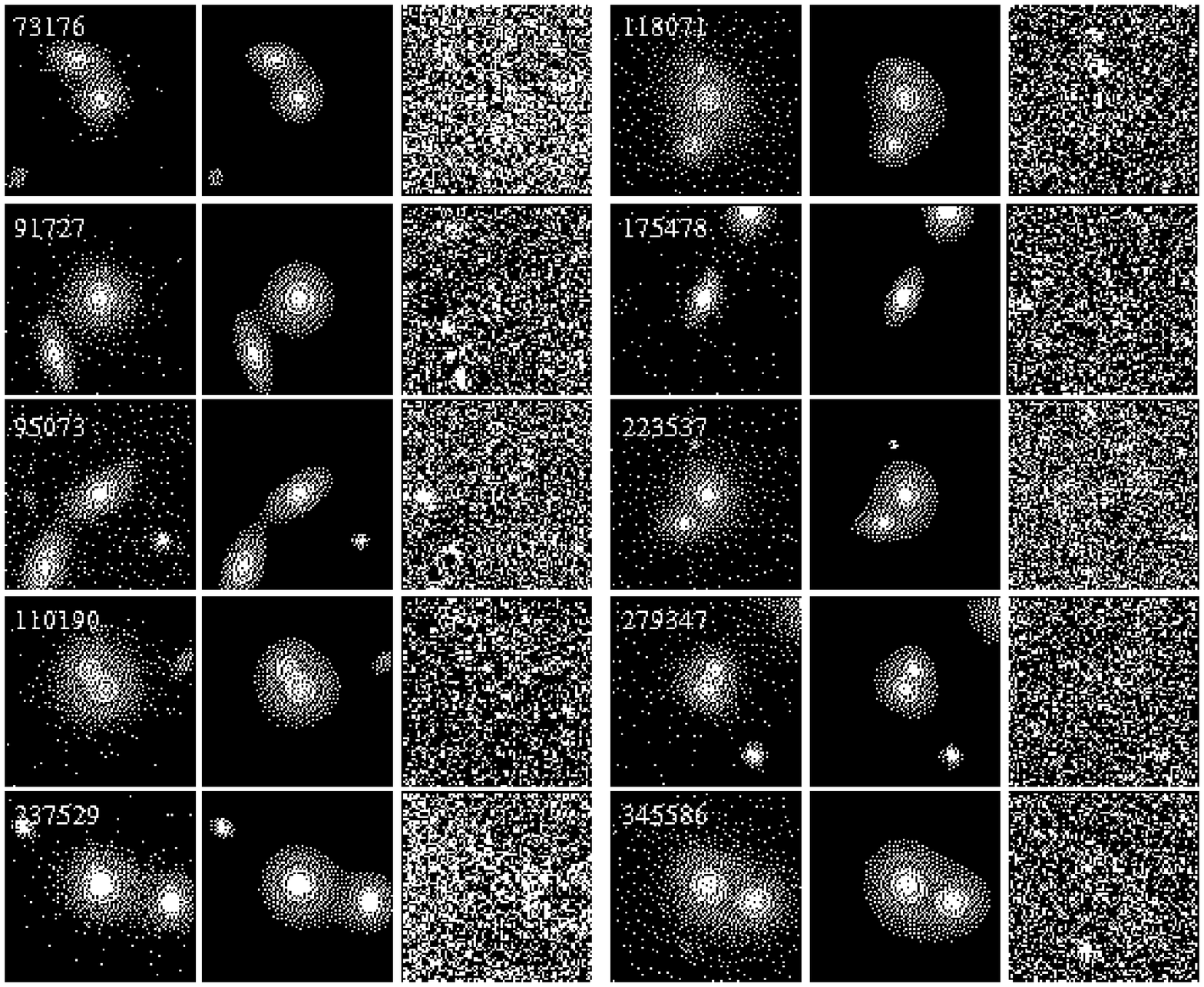}}
\caption[]{
Examples of 10 spec-phot pairs in projection that show no signs of 
disturbance; the central galaxy of each panel has a spectroscopic redshift.
These non-interacting pairs likely have
much larger physical separations than their projected $<30$ kpc separation
suggests, and are either interlopers (separate groups)
or well-separated within a common halo.
All images ($r$-band data, model,
residual with log-scale stretch) are zoomed in to $60\times60$ kpc, and 
are labeled as in Figure \ref{fig:trueproj1}.
\label{fig:nullexs}}
\vspace{-0.2cm}
\end{figure*}

We find 38 pairs of massive galaxies in sampM that we classify as 
major mergers (35 with $d_{12}\leq30$ kpc).
We display the SDSS $r$-band image and corresponding GALFIT residual
of each merger in increasing redshift order (left to right) 
in Figures \ref{fig:dpairs1} and \ref{fig:dpairs2}.
These images leave little doubt that the two galaxies in each pair are in
the midst of merging. We find a variety of strong tidal features including
broad tails (e.g., 311008, 352171, and 274752) such as seen during
the period between second close passage and final coalescence in
dissipationless merger simulations \citep{naab06a} and observations 
\citep[see Fig. 1,][]{bell06a}, and dynamical friction wakes
in the outer stellar envelopes (e.g., 367419 and 258681) as predicted by
\citet{weinberg86} and hinted at in a few BCG systems by \citep{lauer88}.
In addition, we find
bridges (e.g., 301558 and 371303),
plumes (e.g., 150206 and 261132),
diffuse structure (e.g., 294450 and 9993),
and many examples of nonconcentric isophotes
(e.g., 392792, 222852, and 373137), 
which present the strongest indications for tidal contact \citep{lauer88}.
In Figure \ref{fig:nullexs}, we show 10 examples of close (spec-phot) pairs
that have no residual asymmetries and are likely the result of chance
projections. Comparing these non-interacting examples with the 38 mergers in
Figs. \ref{fig:dpairs1} \& \ref{fig:dpairs2} 
clearly demonstrates the fidelity of our 
merger identification scheme.
As the sensitivity of the SDSS
imaging may be too low to detect all interacting pairs of massive galaxies,
our classifications provide a conservative lower limit.
Nonetheless, our sample identifies the strongest cases and serves
as an important dataset for studying the properties of massive
merger progenitors in \S \ref{sec:prog}.

Nearly 70\% (26) of these massive mergers have redshift information
for only one progenitor (spec-phot pairs) as a result of fiber collisions, 
which highlights the importance of our
thorough approach for identifying such systems. 
We estimate that we could
be missing an additional four (11\%) mergers that are photometric-photometric
sources based on the 34\% (26/76) of progenitor galaxies that have only
SDSS photometry.
Quantifying the exact number of massive phot-phot mergers in the DR2-based
group catalog is beyond the scope of this paper. An improved understanding
of the completeness of pairs of merging galaxies in SDSS groups
is one of the aims of our next paper.


\section{Properties of Massive Mergers in Groups and Clusters}

In this section, we explore the properties of the
${\rm M}_{\rm star}\geq10^{11} {\rm M}_{\sun}$ 
mergers that we identified from a
complete sample of $\leq4:1$ mass ratio pairs of massive SDSS
galaxies that we selected from group and cluster-sized halos. We compare
the distributions of basic observables for 
merger pairs and major pairs not classified as mergers,
quantify the nature of the merger progenitors,
make predictions about the remnants, and look for environmental
dependencies in this merging population.

\subsection{Basic Observables}
\label{sec:basic}

\begin{figure*}
\center{\includegraphics[scale=0.8, angle=0]{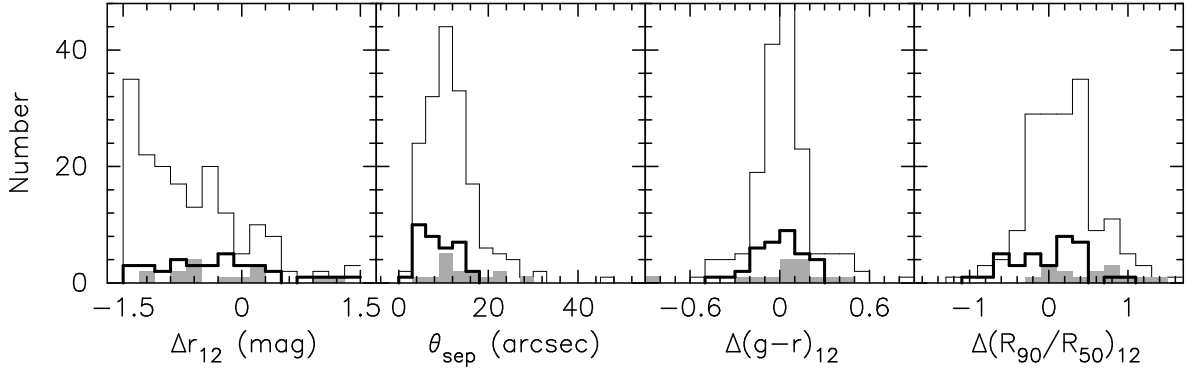}}
\caption[]{Distributions of observables for major ($\leq4:1$ mass ratio)
projected ($\leq30$ kpc) pairs split into three subsets:
merging/interacting (bold lines), known interlopers
(grey bins), and non-interacting systems (thin lines). From left to
right we plot the extinction-corrected $r$-band
Petrosian magnitude difference,
angular separation, extinction-corrected $(g-r)$ Petrosian color difference,
and $r$-band concentration difference. All parameter differences
are defined $\Delta p=p_1-p_2$ such that 1 denotes the galaxy from sampM
and 2 denotes the companion.
\label{fig:pair.obs}}
\vspace{-0.2cm}
\end{figure*}

In Figure \ref{fig:pair.obs}, we plot the distributions of basic observables
that describe each major pair of massive galaxies we selected in 
\S \ref{sec:mgp}. Here we compare the subsets of 35 mergers 
($d_{12}\leq30$ kpc; bold lines),
16 interlopers (i.e., definite non-interacting; grey bins), and
the remaining 170 we classify as non-interacting (thin lines). It is important
to note that a simple selection of major pairs of massive galaxies
in dense environments yields $<16\%$ with obvious signs of merger/interaction.
We find that the pairs that we identify as mergers have some
differences with those without interaction signatures.
Likewise, merging galaxies obviously residing in the same group are
different from projected pairs of galaxies that reside in distinct host halos.

In general, the merger pairs have a flatter $\Delta r_{12}$
distribution than non-interacting pairs, 
with more systems near $\Delta r_{12}=0$, the proxy
for equal-mass mergers, in contrast to the increasing number of
non-mergers towards larger magnitude offsets as expected for
a simple projected pair sample. Nonetheless, there is no statistical
difference between the $\Delta r_{12}$ distributions of mergers
and the subset of known interlopers.
Recall that we select pairs with $\vert \Delta r_{12}\vert \leq1.5$, 
but here we show the $\Delta r_{12}$ distribution to illustrate
that some spec-phot pairs have $\Delta r_{12}>0$; i.e.,
the source without SDSS spectroscopy is more massive.

Merging pairs tend to have smaller angular separations
$\theta_{\rm sep}$
compared with non-interacting pairs and interlopers.
In terms of the colors and concentrations of galaxies in pairs, we
find little difference between the interacting and non-interacting
subsets. Owing to our selection bias for red galaxies 
(see Fig. \ref{fig:cm.sampM})
and the stronger clustering of red galaxies
\citep[e.g.,][]{zehavi02}, it is not surprising that
the color difference $\Delta(g-r)_{12}=(g-r)_1-(g-r)_2$ distributions 
are narrow and peaked near zero. Likewise, given the broad range of
concentrations ($2<R_{90}/R_{50}<4$) found for SDSS Main galaxies 
\citep[e.g.,][]{hogg02}, the relatively small concentration differences
$\Delta(R_{90}/R_{50})_{12}$ are consistent with matched morphologies of
similarly red galaxies.
We note a mild difference between the $\Delta(g-r)_{12}$ and
$\Delta(R_{90}/R_{50})_{12}$ distributions of merging and interloper subsets,
such that the physically unassociated pairs have an increased chance to
be composed of a red massive-group member with a blue, later-type projected
companion.

We check whether or not
any of the basic pair properties in Figure \ref{fig:pair.obs}
depend on the redshift $z_1$ or stellar mass ${\rm M}_1$ of the pair
member from sampM. Only $\theta_{\rm sep}$ depends on $z_1$,
as expected for a sample limited to 30 kpc maximum projected separations.
The different subsets (mergers, interlopers, non-interacting) are 
independent of $z_1$ and ${\rm M}_1$, and hence we conclude that the initial
selection of sampM did not impart biases on our ability to 
classify mergers in a larger sample of major pairs.
Moreover, despite the differences we find between the observables of
merging and non-interacting galaxy pairs, we
cannot distinguish different subsets of major pairs based on these
differences alone.

\begin{figure*}
\center{\includegraphics[scale=0.8, angle=0]{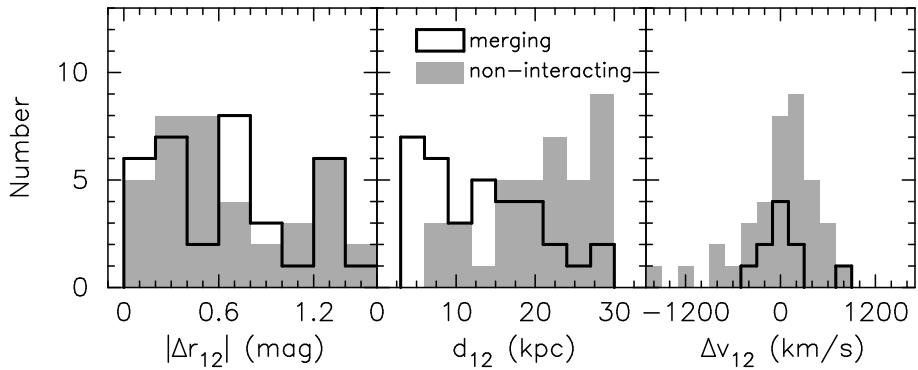}}
\caption[]{Comparison of the properties of 35 major-merging (bold lines) and
39 non-interacting (grey bins) galaxy pairs 
where both galaxies belong to the same host halo.
From left to right, the relative properties of progenitor
galaxies 1 and 2 are absolute value of
the $r$-band magnitude difference,
projected spatial separation in kiloparsecs, and velocity difference.
Only 10 of the 35 major interacting pairs with $d_{12}\leq30$ kpc
have spectroscopic information
for both progenitors to enable the calculation of $\Delta v$.
\label{fig:pair.mergVnon}}
\vspace{-0.2cm}
\end{figure*}

As we mentioned in \S \ref{sec:mgp}, a spectroscopic close
pair of galaxies that belong to the same host dark-matter halo
may reside on opposite sides of the group, and thus have much
larger real space separations than their projected separations imply.
On the other hand, merger pairs by definition must be in close physical
proximity. As such, for pairs where both galaxies are members of the
same group, we compare in Figure \ref{fig:pair.mergVnon} the merging 
and non-interacting subsets
in terms of their projected spatial ($d_{12}$) and 
velocity ($v_{12}$) separations.
We find that the presence/absence
of residual asymmetries clearly produces different $d_{12}$ distributions
consistent with the non-interacting pairs being drawn from a much broader
distribution of real-space separations than the mergers.
Moreover, the declining number of mergers with increasing $d_{12}$ suggests
that wider-separation pairs do not typically include tidal distortions 
that are apparent in the SDSS imaging.
Similarly, we find a more narrow distribution of $v_{12}$ for mergers
compared to the non-interacting pairs in a matched group, but the
significance of this is unclear owing to the large number of 
spec-phot mergers without $z_2$ measurements (25 out of 35).
There is no substantial difference between $\vert \Delta r_{12}\vert$
for the two subsets.

\subsection{Nature of Progenitors}
\label{sec:prog}

In the Introduction, we outlined the importance of improving our
understanding of the progenitors of massive mergers.
Here we use concentration, rest-frame color, and stellar mass to 
explore the properties of the progenitor galaxies in our total sample of
38 mergers; we tabulate information for all 76 progenitors in 
Table \ref{tab:dp}. 

\begin{figure*}
\center{\includegraphics[scale=0.8, angle=0]{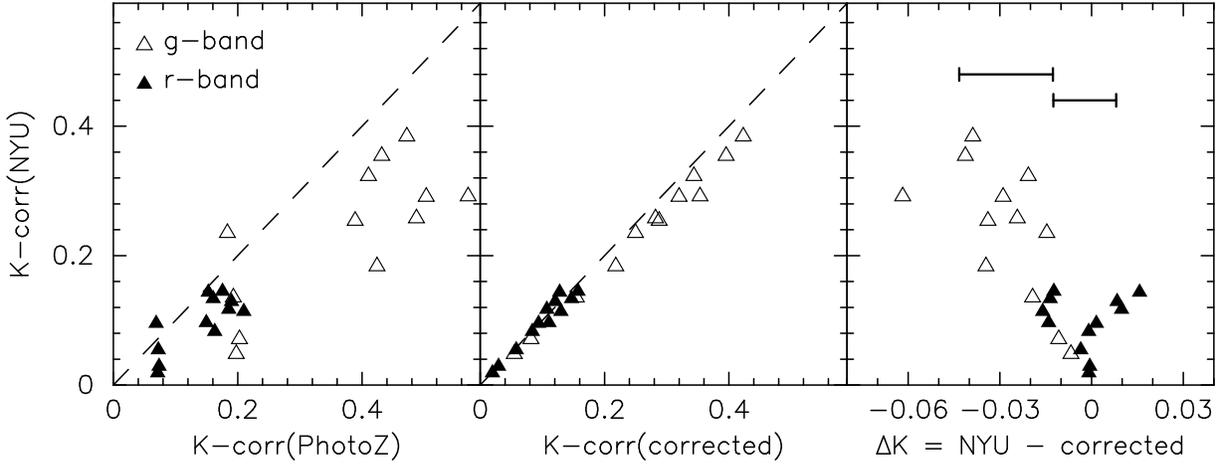}}
\caption[]{$K$-corrections for $g$ (open triangles) and $r$ (solid triangles)
passbands shifted to $z=0$ for the
subset of 12 companion galaxies in major-merger pairs where
spectroscopic redshifts are available for both progenitors.
We plot the accurate NYU\_VAGC $K$-corrections versus those from
PhotoZ (left), corrected using (\ref{eq:2}) (middle), and
the relative difference between the NYU\_VAGC and our corrected values
(right); the error bars show the mean and scatter of the offsets in
each passband.
\label{fig:kcorrs}}
\vspace{-0.2cm}
\end{figure*}

Two thirds of the merger sample have spectroscopic information for
only one of the progenitors as a result of fiber collisions 
(\S \ref{sec:inter}). To obtain rest-frame
quantities for these companions we use $K$-corrections
downloaded from the SDSS PhotoZ table, which we then correct to the redshift
of the merger; i.e., we assume $z_2=z_1$.
For all photometric sources in SDSS, PhotoZ provides photometric 
redshifts $z_{\rm phot}$ and related $K$-corrections $K(z_{\rm phot})$
to shift quantities to $z=0$. For our subset of merger pairs we find that
$z_{\rm phot}$ is systematically larger than $z_1$, and thus
$K(z_{\rm phot})$ is an overestimate.
In the left panel of Figure \ref{fig:kcorrs}, we show the $g$ and $r$-band
$K(z_{\rm phot})$ bias relative to $K(z_2)$
for the 12 mergers in our sample where we have spectroscopic
information for both galaxies. We estimate the correct $K$-correction
for a given passband
\begin{equation}
K(z_2) = K(z_{\rm phot}) \frac{ \log_{10}(1+z_2) }{ \log_{10}(1+z_{\rm phot}) } ,
\label{eq:2}
\end{equation}
by assuming $K(z)\propto2.5\log_{10}(1+z)$ \citep{blanton03b}.
As we demonstrate in the middle and right panels of Figure \ref{fig:kcorrs}, our
method provides excellent $K$-correction estimates with a smaller than 
$\pm0.02$ mag scatter, a -0.03 mag $g$-band offset, and no $r$-band offset.
In this manner, we obtain $^{0.0}(g-r)$ and ${\rm M}_{\rm star}$ estimates
for each photometric progenitor from its extinction-corrected color
downloaded from the SDSS PhotoTag table.

Among the SAT-SAT mergers, there are three spec-phot pairs
(336039, 364190, and 373137) where the 
photometric progenitor 
is more massive than the host group's central (brightest) galaxy. We,
therefore, assume that this galaxy is in fact the CEN and add these pairs
to the CEN-SAT merger subset. Our final sample of 38 pairs of massive
merging galaxies includes 21 CEN-SAT and 17 SAT-SAT systems. 
We distinguish between the CEN-SAT and the SAT-SAT
mergers in the remaining plots. 

\subsubsection{Progenitor Morphology}

\begin{figure*}
\center{\includegraphics[scale=0.8, angle=0]{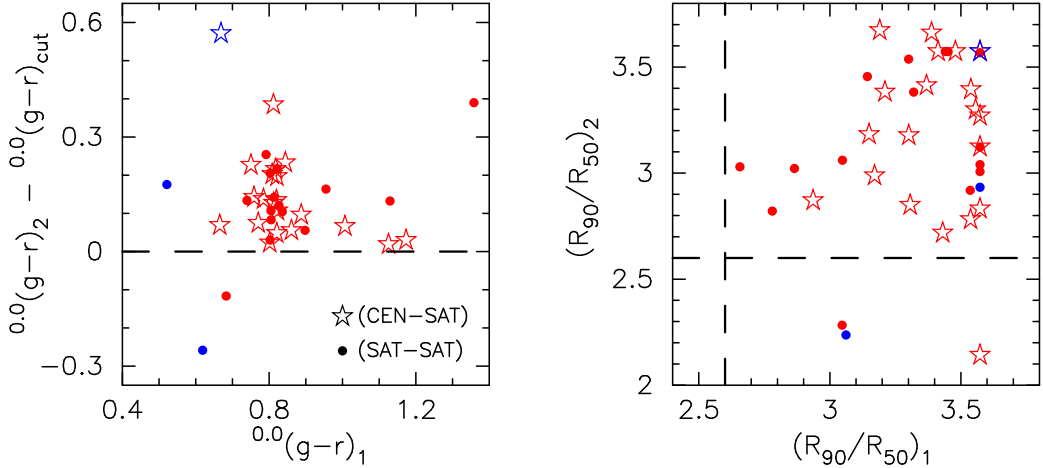}}
\caption[]{Colors and concentrations of the progenitors of 
massive major mergers.
Mergers involving a CEN and SAT galaxy (stars) are 
distinguished from those involving two SATs (circles).
Data points are color-coded to represent blue/red sequence color
of the progenitor in sampM (galaxy number 1).
{\it Left:} relative rest-frame $(g-r)$ color of the companion galaxy (number 2)
with respect to the blue/red cut plotted as a function of progenitor number 1
color. Red points above the dashed line represent red-red mergers.
{\it Right:} $r$-band central-light concentrations of progenitor 2 versus
progenitor 1. Dashed lines show the crude
early/late morphology cut of $R_{90}/R_{50}=2.6$.
\label{fig:prog.type}}
\vspace{-0.2cm}
\end{figure*}

We explore the color and concentration of the progenitor galaxies
in massive merger pairs in Figure \ref{fig:prog.type}. In the left panel,
we plot the rest-frame color of progenitor number 2 relative to the blue/red
sequence boundary shown in Figure \ref{fig:cm.sampM} as a function of the 
color of progenitor number 1 from sampM for each merger pair. 
The data points in both panels of Figure \ref{fig:prog.type} are color-coded
to distinguish blue or red-sequence $^{0.0}(g-r)_1$ colors, and data 
above the dashed line have red $^{0.0}(g-r)_2$ colors. We find that
$90\pm5\%$ of the massive mergers we identify are comprised of two red progenitors;
only one merger is blue-blue and three are mixed pairs.
In the right panel, we show the central-light concentration
of progenitor 2 plotted against that of progenitor 1.
Consistent with the high fraction of red-red mergers, $92\pm4\%$
of the mergers are comprised of two concentrated progenitors with
$R_{90}/R_{50}>2.6$, the fiducial value for early-type morphologies
(see \S \ref{sec:mgp}). Three mergers are made up of an early/late mix
according to concentration, with one of each red-red, red-blue, and blue-blue.

The nature of the progenitors appears to depend little on whether
the merger is positioned at the center of the host group or
is between a pair of SAT galaxies. Owing to the small-number statistics, the 
slight decrease in the red-red
merger fractions from 95\% (CEN-SAT) to 82\% (SAT-SAT), and likewise for
early-early mergers from 95\% (CEN-SAT) to 88\% (SAT-SAT), 
are consistent with no difference.
Generally speaking, the major mergers that will produce 
${\rm M}_{\rm star}>10^{11} {\rm M}_{\sun}$ remnants in massive groups are
between two red-sequence spheroids that have little cold gas for
star formation and are presumably dissipationless.
The properties of these low-redshift mergers match the six $0.1<z<0.9$ 
dissipationless mergers in \citet{bell06a}.

\subsubsection{Progenitor Mass Ratios}
The major mergers we have identified are drawn from pairs with
$\vert \Delta r_{12}\vert \leq1.5$ mag, our proxy for 4:1 to 1:1
mass ratios (\S \ref{sec:mgp}). Here, we explore the actual
stellar mass ratios of the merger progenitors. Overall, the
Petrosian color-derived ${\rm M}_{\rm star}$ estimates for sampM
are well-behaved as demonstrated by the tight red-sequence of CEN
and SAT members in Figure \ref{fig:cm.sampM}. We note, however, that
there are a handful of extreme outliers in color-mass space such
that some massive group galaxies have very red colors, especially
at the high-mass tip of the red sequence. Large systematic errors in color
translate into errors in ${\rm M}_{\rm star}$, which is a critical
issue when trying to ascertain the progenitor mass ratios. Nevertheless,
it is unclear whether the measured colors are the result of an error
in the photometric pipeline or simply the intrinsic nature of a rare population.

\begin{figure*}
\center{\includegraphics[scale=0.8, angle=0]{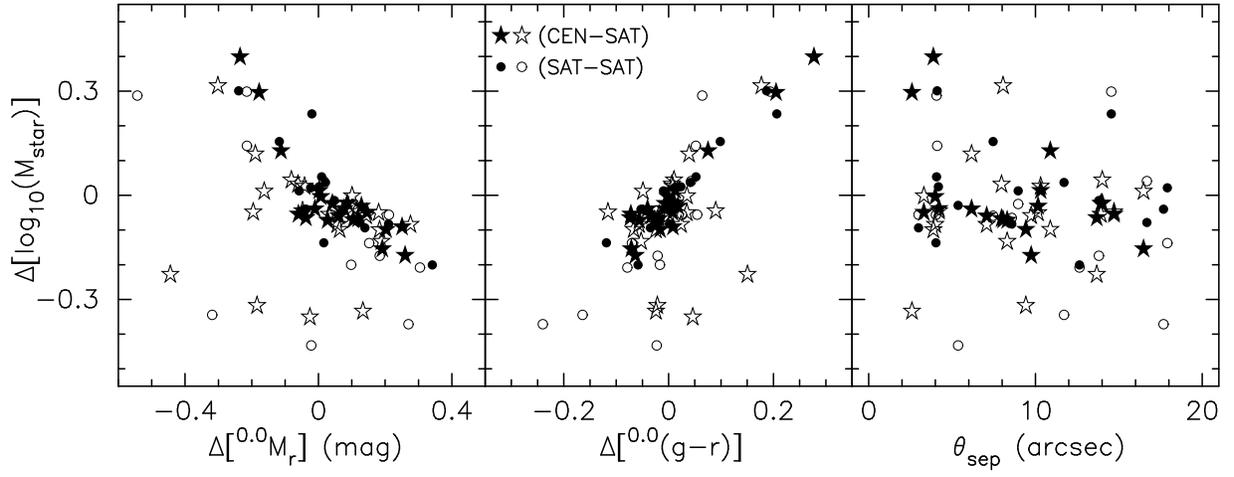}}
\caption[]{The systematic stellar mass uncertainty for progenitors of the 38
massive mergers plotted against the corresponding uncertainty in
absolute $r$-band magnitude, rest-frame $(g-r)$ color, and pair
separation. All relative differences between
quantities based on SDSS Petrosian and Model magnitudes are
such that $\Delta = {\rm Petrosian} - {\rm Model}$. Solid and open symbols
represent galaxy number 1 and 2, respectively.
\label{fig:mass_err}}
\vspace{-0.2cm}
\end{figure*}

We attempt to quantify the amplitude of systematic uncertainties in our
stellar mass estimates from issues related to the SDSS photometry
by recomputing ${\rm M}_{\rm star}$ for all 76 merger progenitors
using SDSS Model\footnote{In addition to standard Petrosian magnitudes,
the SDSS photometry includes measures of galaxy 
flux from the best-fit model, either a de Vaucouleurs or an exponential,
to the $r$-band image profile.}
magnitudes in place of Petrosian quantities
in (\ref{eq:1}). In Figure \ref{fig:mass_err}, we plot the relative 
difference between the masses derived with each type of magnitude as a
function of the $^{0.0}M_r$ and $^{0.0}(g-r)$ differences (Petrosian-Model),
and $\theta_{\rm sep}$. We find that the bulk (75\%) of the progenitors
have a small ($\leq0.15$ dex) but systematic shift towards lower
masses, which correlates with fainter $^{0.0}M_r$, when using Model magnitudes.
This subset has a tight locus of $\Delta[^{0.0}(g-r)]$ comparable to the quoted
$\sim0.04$ mag random error for Petrosian colors. The remaining 25\%
of the progenitors have systematic color offsets as large as 
$+0.30(-0.25)$ mag resulting in a greater than factor of 2 shift
in ${\rm M}_{\rm star}$,
or more than twice as much as the expected 0.10-0.15 dex 
systematic uncertainty (\S \ref{sec:mgp}).
In Table \ref{tab:dp}, we note the progenitors with $\geq0.3$ dex difference
between their Petrosian and Model-based ${\rm M}_{\rm star}$ estimates.
We find no dependence of these mass offsets on CEN versus SAT, nor on the
angular separation of the pairs.
One possible explanation for the large photometric variances could be related to
known pipeline errors for very close pairs \citep{masjedi06},
yet very few of our projected pair sample have $\theta_{\rm sep}<3\arcsec$
and it is difficult to understand how close pairs would have boosted flux
in one passband ($r$) but not another ($g$) to account for the very red colors.

\begin{figure*}
\center{\includegraphics[scale=1.1, angle=0]{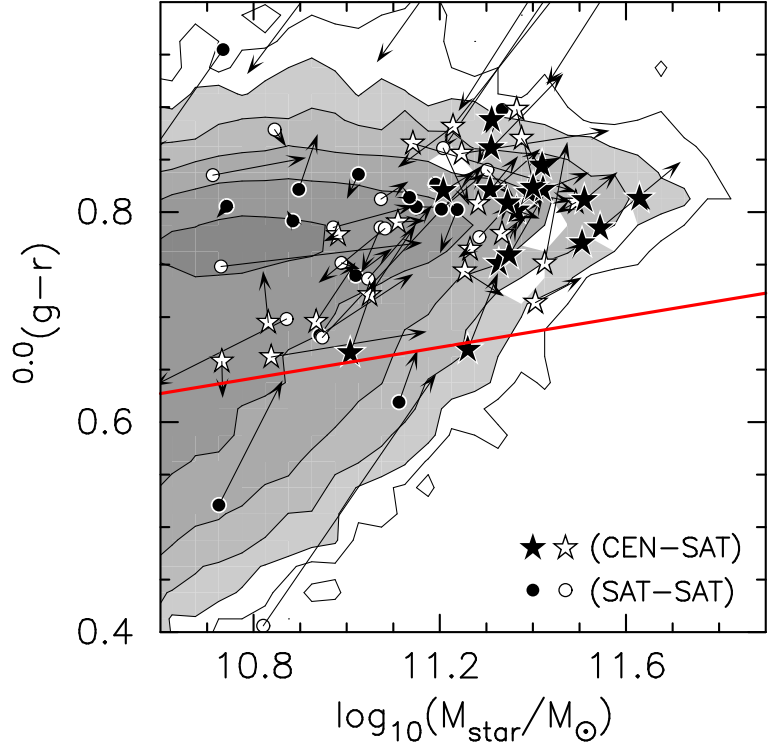}}
\caption[]{
Distribution of the 76 massive merger progenitors in color versus stellar mass.
Symbols distinguish CEN-SAT (stars) and SAT-SAT (circles) mergers with
open symbols representing the progenitor from sampM.
Data points are based on Petrosian photometry with arrows showing the
offset to $^{0.0}(g-r)$ and ${\rm M}_{\rm star}$ values using SDSS Model
magnitudes (see text for details).
The contours and blue/red galaxy division are as in  Fig. \ref{fig:cm.sampM}.
\label{fig:cm.prog}}
\vspace{-0.2cm}
\end{figure*}

We plot the color-stellar-mass distribution of the 76 progenitors
in Figure \ref{fig:cm.prog} using symbols to represent the Petrosian-derived
values and arrows to explicitly show the direction and amplitude of the shifts
to Model-derived values in this parameter space. We note that the extremely-red
outliers in Petrosian space have Model colors more in accord with normal
red galaxies. Generally speaking,
most of the progenitors occupy the massive end of the red 
sequence; 80\% have ${\rm M}_{\rm star}>10^{11} {\rm M}_{\sun}$.
This means that some very massive galaxies continue to be assembled
in the low-redshift universe. Yet, as a result of selection effects, we
cannot determine the significance of the small number of progenitors with
${\rm M}_{\rm star}<10^{11} {\rm M}_{\sun}$.
The selection of sampM creates
a bias insofar as the percent contribution of massive halo members to
the overall DR2 galaxy population in the $z\leq0.12$ volume decreases 
significantly as a function of stellar mass (see Table \ref{mstar_counts}).
In other words, our halo mass-limited selection misses vast numbers of
galaxies with ${\rm M}_{\rm star}<10^{11} {\rm M}_{\sun}$ simply because
they live in halos with ${\rm M}_{\rm halo}<2.5\times10^{13} {\rm M}_{\sun}$.
The importance of massive major mergers in lower-mass halos will be
the subject of a followup paper.

\begin{figure*}
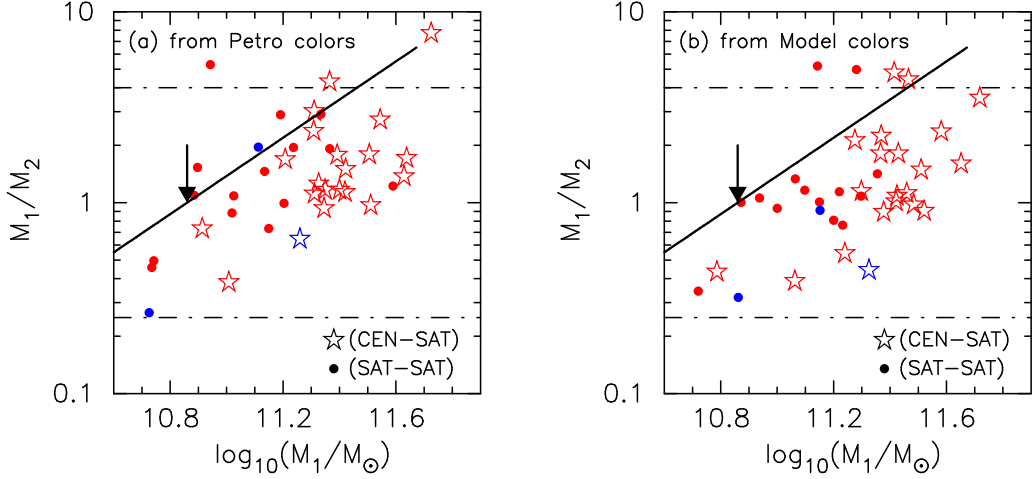

\centering
\mbox{
  \includegraphics[scale=0.8, angle=0]{mcintosh_fig15.eps}
  \hspace{0.75cm}
  \includegraphics[scale=0.8, angle=0]{mcintosh_fig16.eps}
}
\caption[]{Stellar mass ratios of the progenitors of massive major mergers
plotted as a function of the stellar mass of progenitor number 1. 
The two panels show results for color-derived stellar masses based 
on Petrosian ({\it left}) and Model ({\it right}) SDSS magnitudes.
Symbols and color coding are as in Fig. \ref{fig:prog.type}. The
dot-dashed lines show the 4:1 mass ratio boundary of major mergers.
Mergers with ${\rm M}_1/{\rm M}_2<1$ have no redshift for
the more-massive primary galaxy (i.e., are spec-phot pairs).
Mergers on the solid diagonal line have one progenitor with a mass equal to
${\rm M}^{\ast}=7.24\times10^{10} {\rm M}_{\sun}$ (marked by the arrow).
\label{fig:prog.mass}}
\vspace{-0.2cm}
\end{figure*}

In Figure \ref{fig:prog.mass}, we show the stellar mass ratios
of the progenitors in our
sample of mergers as a function of the mass ${\rm M}_1$ of the progenitor
drawn from sampM. There is little qualitative difference between
the distributions of ${\rm M}_1/{\rm M}_2$ based on Petrosian or Model
photometry. 
Both CEN-SAT and SAT-SAT mergers have mass ratios mostly between
2:1 and 1:1,
with the primary progenitors in central mergers tending toward
higher masses than those in SAT-SAT mergers. We discuss the implications
of these mass ratios on the merger assembly of massive galaxies in 
\S \ref{sec:Disc1}.

\subsubsection{The Predicted Color-Mass Distribution of Massive Remnants}
\label{sec:cmplane}

Recall that besides the sample of 38 merger pairs we have also identified
seven massive mergers based on their disturbed morphologies
(\S \ref{sec:mgp}; Fig. \ref{fig:mergers}). Under the assumption that
these morphologically-identified mergers
are examples of an advanced evolutionary stage between interacting pairs
of massive galaxies
and the final coalesced remnant\footnote{This is a fair assumption given
that all seven morphologically-disturbed
mergers have stellar masses in excess of $10^{11} {\rm M}_{\sun}$.}, 
it is worthwhile to compare their positions in the color-stellar-mass
plane with the predicted locations for the remnants of the merger pairs.
For each pair of progenitors we calculate the remnant's
final mass ${\rm M}_{\rm rem}={\rm M}_{\rm p}+f{\rm M}_{\rm s}$
and its mass-weighted color 
\begin{equation}
(g-r)_{\rm rem} = -2.5\log_{10}\left[ \frac{ {\rm M}_{\rm p} } { {\rm M}_{\rm rem} }10^{-0.4(g-r)_{\rm p}} + \frac{ f{\rm M}_{\rm s} }{ {\rm M}_{\rm rem} }10^{-0.4(g-r)_{\rm s}} \right] ,
\label{eq:3}
\end{equation}
where the primary progenitor is more massive than the secondary
by definition (i.e., ${\rm M}_{\rm p}\geq{\rm M}_{\rm s}$).
The factor $f$ allows us to adjust the fraction of the secondary
progenitor's initial mass that is included in ${\rm M}_{\rm rem}$.
In what follows we use Petrosian-based quantities.

\begin{figure*}
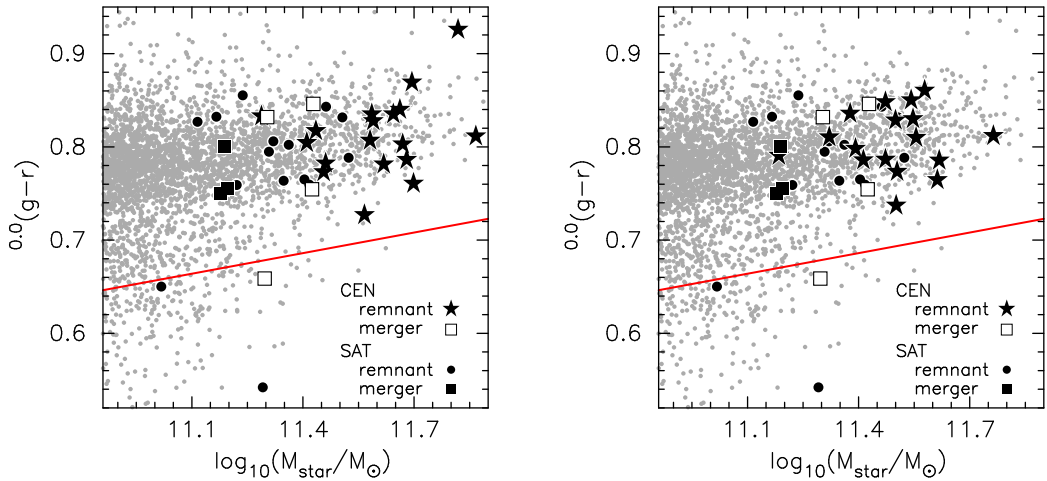

\centering
\mbox{
  \includegraphics[scale=0.7, angle=0]{mcintosh_fig17.eps}
  \hspace{0.75cm}
  \includegraphics[scale=0.7, angle=0]{mcintosh_fig18.eps}
}
\caption[]{
Predicted stellar masses and mass-weighted colors of massive merger remnants 
compared with observations of 
disturbed-morphology mergers presumed to be nearing
final coalescence. Small grey circles show all sampM galaxies
more massive than
${\rm M}^{\ast}=7.24\times10^{10} {\rm M}_{\sun}$.
Open (CEN) and filled (SAT) squares represent the
seven mergers shown in Fig. \ref{fig:mergers}; stars (from CEN-SAT mergers) and
circles (from SAT-SAT mergers) represent the predicted remnants of the 
38 merger pairs.
{\it Left panel:} the simple 
assumption that all of the mass from both progenitors is added to
the final remnant.
{\it Right panel:} the assumption that 50\% of the SAT progenitor mass is
added to the ICL if the merger is at the group center.
The blue/red galaxy division is as in  Fig. \ref{fig:cm.sampM}.
All data are based on Petrosian quantities.
\label{fig:cm.remn}}
\vspace{-0.2cm}
\end{figure*}

In each panel of Figure \ref{fig:cm.remn} we show the seven
morphologically-identified mergers as squares with CEN (SAT) examples
distinguished by open (closed) symbols. Six of the seven have red-sequence
colors, and all CENs are more massive than the SATs.
We first compare the color-mass distribution of these mergers with
the predict distribution of remnants from the 38 merger pairs
under the simple assumption that the total mass of the secondary
is always accreted onto the remnant ($f=1$). 
Nearly all remnants have red-sequence
colors reflecting the nature of their progenitors. 
In terms of the stellar masses of the
observed mergers compared to the predicted remnants, 
we find better agreement with SATs than for CENs.
Three quarters of the future remnants at the centers of massive
groups are more massive than the four morphologically-identified CEN mergers.
With small number statistics it is difficult to make definitive comparisons.
It is possible that the time interval that a late-stage merger 
is apparent depends
on stellar mass, such that higher-mass mergers coalesce into a single
object faster. Another possibility is
that some mass is lost during the merging process.

\citet{zibetti05} found that the intracluster light (ICL) within 100 kpc of the
group or cluster center makes up as much as
40\% of the total cluster luminosity (galaxies$+$ICL),
and they showed that the stars making up the ICL have the same colors
as the old-stellar light from the massive galaxies in the
intracluster environment. Therefore, it is conceivable that some stellar mass
from the massive, red, CEN-SAT mergers deep in the potential wells of 
large groups and clusters winds up in the ICL rather than as part of the
central remnant galaxy. Various groups have argued that disruption
of SAT galaxies through tidal stripping and heating
can remove 10--80\% of their stellar mass and account for
the ICL \citep{monaco06,white07,conroy07c}.
These theories provide a way
to reconcile the predicted merger-driven mass growth above
$10^{11} {\rm M}_{\sun}$ in a $\Lambda$CDM cosmology with the little growth
that is observed in the stellar mass function \citep[e.g.,][]{wake06,brown07}.
In the right panel of Figure \ref{fig:cm.remn}, we try a highly
conservative test of the latter scenario by assuming $f=0.5$ for CEN-SAT
remnants, but $f=1$ for SAT-SAT mergers. This assumption implies that each
massive SAT merging with the center of its host potential well
would lose 50\% of its present stellar mass by the time it coalesced from
an average projected group-centric distance of 15 kpc. 
In the previous section,
we show that these CEN-SAT mergers have mass ratios typically within a factor 
of two of unity, and these systems are clearly separated by distances much less
than the ICL half-light radius. These facts suggest either (i) a much lower SAT
mass loss than our conservative assumption, or (ii) the SAT masses at the ICL
half-light radius were in excess of the CEN with which they will eventually
merge. Another possibility is that the CEN-SAT masses are much more disparate
as a result of the standard SDSS photometry systematically underestimating the
CEN luminosities \citep{lauer07a}. Resolving these issues is beyond the
scope of this paper. Instead we simply point out that
in terms of relative stellar mass from SDSS photometry, we see
better agreement between the observed mergers and the predicted remnants
at the centers of massive groups if we assume that only half of the SAT mass
ends up in the remnant, which suggests 
that major mergers at the bottom of the potential well in groups and
clusters could be an important source for the ICL.

\subsection{Environmental Dependencies}
\label{sec:enviro}

One of the key goals of our study is to quantify the environmental
dependencies, if any, of massive mergers. Here we use the host's halo mass,
and the distinction between CEN (brightest) and SAT members,
to explore the environments of the mergers that we have identified in
large SDSS groups and clusters from the local universe. 
In what follows, we consider
the combined sample of 45 massive mergers: 38 close pairs identified by residual
asymmetric structure plus seven single sources identified by their
morphologically-disturbed appearance.

\subsubsection{Preference for Central Merging}
\label{sec:cenmerging}

We find that the centers of massive groups and clusters appear to
be the preferred environment for the major-merger assembly of
present-day ${\rm M}_{\rm star}>10^{11}{\rm M}_{\sun}$ galaxies. 
More than half of the
mergers we identify involve the central (most-luminous) member of the 
host dark-matter halo, yet there are five times less CENs than SATs
to merge with in sampM. Thus, on average, 3\% of massive groups with
$z\leq0.12$ have a major merger, but less than 1\% of all massive
galaxies within these groups are merging.

\begin{figure*}
\center{\includegraphics[scale=0.8, angle=0]{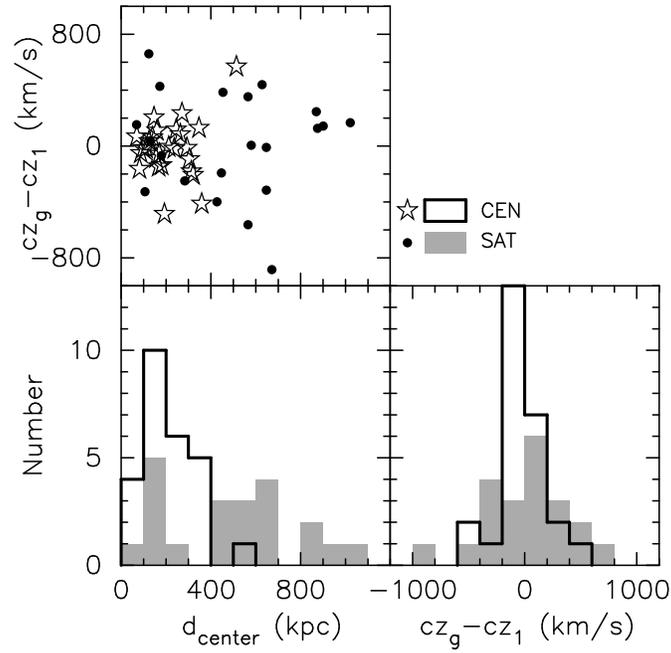}}
\caption[]{Comparison of the group-centric properties of major mergers
occurring at group centers (bold lines, stars) or between two
SAT galaxies (grey bins, circles).
The {\it top} panel shows the transverse projected offset in kpc
versus the radial offset in km/s of each merger relative
to the luminosity-weighted group center. The {\it bottom} panels
provide the separate group-centric property distributions.
\label{fig:merg.cenVsat}}
\vspace{-0.2cm}
\end{figure*}

In Figure \ref{fig:merg.cenVsat}, we compare the group-centric properties
of the CEN and SAT mergers.
We find that mergers involving a CEN are significantly closer to the
luminosity-weighted center of their host group than mergers between
SAT galaxies. The average projected group-centric distance
of CEN mergers is 210 kpc, compared to 490 kpc
for SATs. Moreover, relative to the luminosity-weighted group redshifts,
the CEN mergers have a narrower distribution of velocity offsets
($\sigma=200$ \kms) than the SAT mergers ($\sigma=370$ \kms). 
The small group-centric offsets
of the CEN mergers are consistent with them residing at the
bottom of their halo's potential well, where dynamical friction
is maximum. In contrast, most merging SATs have large 
group-centric offsets as expected given their rank within their host group.
At face value, these results indicate that mergers between massive
SATs do occur, yet in terms of their
morphologies (Fig.\ref{fig:prog.type}) 
and mass ratios (Fig.\ref{fig:prog.mass})
there are no clear differences between CEN-SAT and SAT-SAT merger progenitors. 

A merger between two massive galaxies likely occurs at the dynamical
center of a common dark-matter halo. If the merger is between two
SATs, then we may be witnessing a merger at the center of a subhalo
that merged with the larger host halo\footnote{The
halo-based group finder of \citet{yang05a} used to produce the SDSS
group catalog does not have the ability to distinguish subhalos within
the halo defining each galaxy group.}.
Another possibility is that the SAT merger represents the true dynamical
center of the host halo. Indeed, we find that 20\% (4/20) of the SAT
mergers reside closer to the center of the group's projected galaxy distribution
than the spectroscopic CEN galaxy identified by the group catalog, 
and have a total stellar mass estimate ${\rm M}_1+{\rm M}_2$ that is
greater than the mass of the CEN (${\rm M}_{\rm CEN}$).
We identify these four pairs
in Table \ref{tab:dp} and explore their inclusion/exclusion in the
CEN-SAT merger subset in our analyzes
of central merger frequencies and mass accretion rates in the following
sections.
 
It is also possible that a significant fraction of the SAT mergers are
at the center of a distinct halo seen in projection along
the line of sight to the host halo. This explanation would explain the 
group-centric differences in Figure \ref{fig:merg.cenVsat}, and the similarities
in color, concentration, and mass ratios that we observe.
We note that 6/20 SAT mergers have ${\rm M}_1+{\rm M}_2\geq{\rm M}_{\rm CEN}$
and large projected group-centric distances, providing circumstantial
evidence for membership in a separate group from the host of the CEN galaxy.
Yet, a simple calculation shows that there is only a 10\% chance for a
line-of-sight projection of a distinct group with
${\rm M}_{\rm halo}\geq10^{13}{\rm M}_{\sun}$ within 1 Mpc radius and
$\pm400$ \kms\ depth (following the
group-centric properties of the SAT-SAT mergers in
Fig. \ref{fig:merg.cenVsat}). This estimate is an upper limit based on the
mean number density of groups that
typically host a $10^{11}{\rm M}_{\sun}$ CEN galaxy
\citep[$10^{-3.5}$ Mpc$^{-3}$,][]{mo02}, and the assumption that the correlation
strength between groups increases the local density relative to the
mean by a factor of 10.
Therefore, our observed frequency of 3\% of groups with central mergers
implies that we should find only three SAT-SAT mergers that are misidentified
CEN-SAT systems from a projected group. We find 16--20 SAT-SAT mergers
in 845 groups (1.9--2.4\% depending on whether we consider the four
mentioned above to be
at the center of their host), indicating that most are correctly identified
as SAT-SAT interactions. Given the large velocity dispersions of high-density
environments, true SAT-SAT mergers are not expected. While we do find
large group-centric velocity offsets for SAT-SAT mergers 
(Fig. \ref{fig:merg.cenVsat}), for the subset of 12 spec-spec mergers
we find no significant difference between
the small velocity separations ($v_{12}$, see Fig. \ref{fig:pair.mergVnon})
of CEN-SAT and SAT-SAT mergers. Therefore,
we tentatively conclude that 
massive SAT-SAT mergers identify the centers 
of large subhalos that have recently accreted onto their host.

\subsubsection{Merging Dependence on Halo Mass}
\label{sec:halomassdep}

\begin{figure*}
\center{\includegraphics[scale=0.8, angle=0]{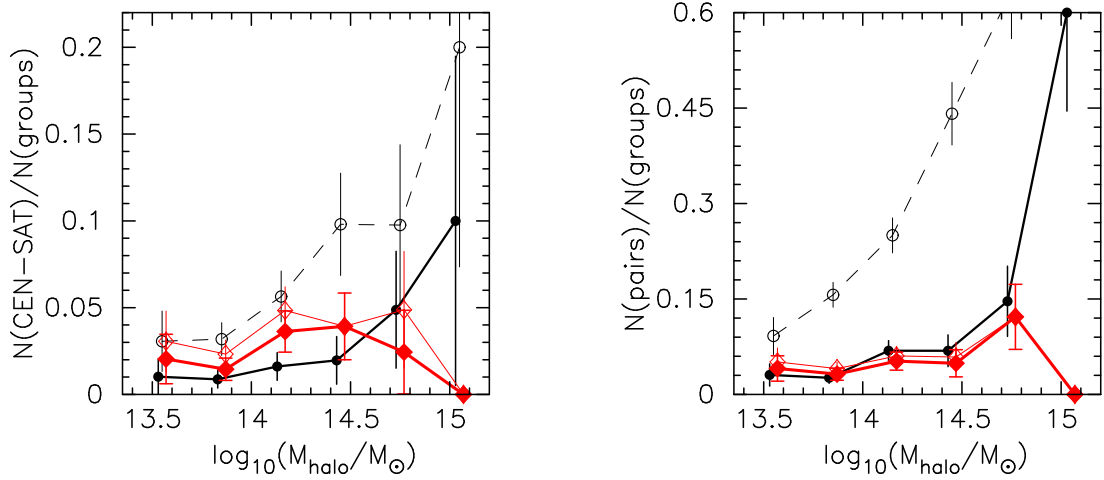}}
\caption[]{Halo-mass dependence for the frequency of major pairs of galaxies
with ${\rm M}_1+{\rm M}_2\geq10^{11} {\rm M}_{\sun}$ in groups with
${\rm M}_{\rm halo}>2.5\times10^{13}{\rm M}_{\sun}$.
The fraction of groups with CEN-SAT (left panel)
and combined CEN-SAT plus SAT-SAT (right panel) projected pairs
are plotted as a function of halo mass in 0.3 dex bins.
The dashed line with open circles is for all pairs with $d_{12}\leq30$ kpc, 
the solid black line with filled circles is the subset of
close pairs that reside in the same host halo (spec-spec only), 
and the bold red line with
filled diamonds denotes the subset of galaxy-galaxy mergers
(both spec-spec and spec-phot pairs) identified by
our profile fitting method (\S \ref{sec:inter}).
The thin red line with open diamonds in each panel combines the galaxy-galaxy
mergers and the additional mergers identified by highly-disturbed morphologies,
which provides an upper limit to the number of mergers per group that are
detectable in SDSS data. For CEN merging (left panel), this upper limit
includes the addition of four SAT-SAT mergers that appear to be at the
actual dynamical center of their host halo (see \S \ref{sec:cenmerging}).
Poisson errors are shown.
\label{fig:mhalo_dep}}
\vspace{-0.2cm}
\end{figure*}

By identifying the massive galaxy mergers in a
halo mass-limited selection of large groups, we
can for the first time constrain their importance
as a function of halo mass. In the left panel of Figure \ref{fig:mhalo_dep},
we plot the halo-mass dependence for the frequency of groups that have
merger-driven assembly of ${\rm M}_{\rm star}\geq10^{11} {\rm M}_{\sun}$
galaxies restricted to group centers.
We find that the fraction of groups that have a massive merger
at their center (bold red line) is
statistically constant at 3\% over the interval 
$13.4<\log_{10}{({\rm M}_{\rm halo}/{\rm M}_{\sun})}<14.9$; we note that
including the four morphologically-identified CEN mergers plus the four
misclassified SAT-SAT mergers at their host's dynamical center results
in a minor increase to this frequency (thin red line, open diamonds).
We contrast our estimate for the merger frequency
dependence on halo mass, based
solely on galaxies exhibiting obvious tidal features,
to that obtained from simple close projected pairs of massive galaxies
(dashed black line), which grows steadily with halo mass as a result of
the increased projected number density of massive galaxies in dense
environments. The increased chance of projection with increasing 
${\rm M}_{\rm halo}$ also occurs for the subset of spec-spec pairs
that are found
in the same host group (solid black line), but the amplitude
is diminished owing to the high spectroscopic incompleteness in the pairs
that we study. At $\log_{10}{({\rm M}_{\rm halo}/{\rm M}_{\sun})}<14.6$, the
number of spec-spec projected pairs in matched halos is less than that of
all mergers, which by definition must reside in the same host halo,
because the latter include spec-phot pairs.

In the right panel of Figure \ref{fig:mhalo_dep}, we repeat our
analysis of merger and projected pair frequencies as a function of
halo mass using the combined CEN-SAT plus SAT-SAT sample.
When considering all possible mergers that will produce high-mass 
remnants (red lines), 
the frequency is roughly constant at 5\% for 
$13.4<\log_{10}{({\rm M}_{\rm halo}/{\rm M}_{\sun})}<14.9$.
Including/excluding the seven non-pair mergers (morphologically-identified)
does not change these frequencies significantly.
The issues with using simple projected pair
statistics (black lines) to estimate merger frequencies grow rapidly out of hand
for massive groups that contain large numbers of 
${\rm M}_{\rm star}\geq5\times10^{10}{\rm M}_{\sun}$ 
galaxies; e.g., on average nearly half of all
$\log_{10}{({\rm M}_{\rm halo}/{\rm M}_{\sun})}=14.5$ groups have one
major pair of massive galaxies that appear close in projection.

\begin{figure*}
\center{\includegraphics[scale=0.8, angle=0]{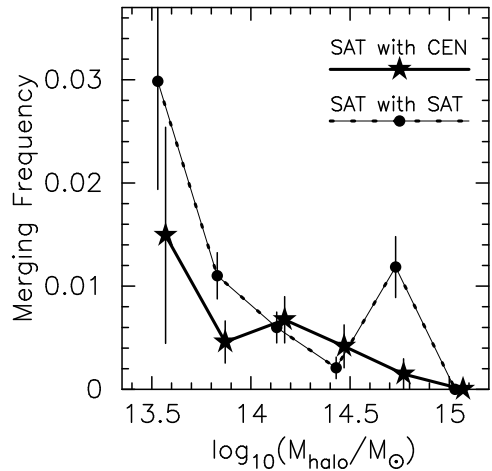}}
\caption[]{Frequency of massive 
(${\rm M}_{\rm star}\geq5\times10^{10}{\rm M}_{\sun}$)
SATs that are involved in merging with either a CEN (solid line with stars)
or another massive SAT (dashed line with circles) galaxy,
as a function of halo mass.
Poisson errors are shown.
\label{fig:mhalo_freq}}
\vspace{-0.2cm}
\end{figure*}

Besides merger frequency per group we can approach massive mergers
from a different perspective and calculate the frequency of
${\rm M}_{\rm star}\geq5\times10^{10}{\rm M}_{\sun}$ SATs that are currently 
involved in a merger that can be identified as such with 
the technique that we use here. We calculate separate frequencies
for merging with a CEN or another massive SAT galaxy in bins of halo mass,
and plot them in Figure \ref{fig:mhalo_freq}. 
As a function of ${\rm M}_{\rm halo}$,
the SAT merging frequencies decrease from a few percent
for our lowest-mass groups, to $\leq1\%$ for groups
larger than ${\rm M}_{\rm halo}=5\times10^{13}{\rm M}_{\sun}$.
The CEN-SAT merging follows a very similar decreasing frequency trend with
increasing ${\rm M}_{\rm halo}$ as SAT-SAT mergers, which is
qualitatively consistent
with dynamical friction and provides more circumstantial
evidence that SAT-SAT mergers are occurring at the dynamical centers
of recently accreted subhalos. Even though SATs have a large relative
velocity dispersion, they can still merger through dynamical friction
if they are both members of a subhalo.

\section{Discussion}
\label{sec:Disc}

We find the first direct observational evidence for an important
population of galaxy-galaxy mergers with total stellar masses
above $10^{11} {\rm M}_{\sun}$ in the local universe. These objects
provide an unprecedented census of the progenitor
properties for the merger-driven assembly of high-mass galaxies, which
we compare to recent predictions from numerical models of galaxy
formation and evolution. Moreover, the existence of these mergers
prove that a measurable amount of stellar mass growth continues
in the massive galaxy population at present times, 
and we compare estimates based on
this sample with other estimates in the literature. Finally,
we have identified mergers restricted to reside in large SDSS groups
and clusters with $z\leq0.12$, thus allowing the first constraints
on the halo-mass dependencies of recent massive merger activity.
While it is well-established that massive
galaxies are more common in such high-density environments, we are missing
much more than 50\% of the population with 
${\rm M}_{\rm star}<4\times10^{11} {\rm M}_{\sun}$ in the local volume,
as Table \ref{mstar_counts} shows. Therefore, we must keep
this caveat in mind when interpreting the conditions for which our
results hold.
In an upcoming study, we are examining the role of
major mergers as a function of stellar mass over the full range of
environments hosting galaxies more massive than 
${\rm M}_{\rm star}=5\times10^{10}{\rm M}_{\sun}$.

\subsection{Massive Merger Progenitors: Observations Meet Theories}
\label{sec:Disc1}

Establishing the luminosity dependence of elliptical (E)
galaxy properties \citep{davies83,bender88,bender92}
set the stage for theories regarding the types of merger progenitors that
would produce the characteristics of low and high-mass 
early-type galaxies (ETGs)\footnote{The distinction between elliptical and
early-type galaxies is often blurred in the literature. We consider Es 
to be a morphological subset of ETGs, which are concentrated and 
spheroid-dominated systems including Es, lenticulars (S0s), and Sa spirals.
When referencing other authors we remain faithful to their choice of
nomenclature.}
galaxies \citep{bender92,kormendy96,faber97}.
We concentrate on modern numerical simulations and semi-analytic models
that attempt to reproduce the kinematic, photometric, and structural
properties observed in massive Es through major merging
\citep{naab99,naab03,khochfar03,khochfar05,naab06a,boylan06,kang07}.
For this discussion
we make the straight-forward assumption that the major mergers that we
have identified will produce remnants that are {\it not unlike} the
${\rm M}_{\rm star}>10^{11} {\rm M}_{\sun}$ galaxy population 
already in place. We can only guess at remnant properties 
(see Fig. \ref{fig:cm.remn}), but in general, massive galaxies on the
red-sequence are typically early-type. 

As we show in Figure \ref{fig:prog.mass}, the progenitor masses are
comparable for the most part, and quantitatively consistent with
the LRG-LRG merger mass spectrum from \citet{masjedi07} under the
assumption that companions merge on dynamical friction time scales. 
$N$-body simulations \citep[e.g.][]{naab99} have long shown
that ${\rm M}_1/{\rm M}_2\approx1$ are necessary to produce the lack of
significant rotation observed in massive Es. Yet, a near unity mass ratio
alone is not sufficient to produce the predominance of boxy and anisotropic
Es found at high luminosity \citep{naab03,naab06a}.
To match the decreasing fraction of rotational support and increasing
fraction of boxiness in more luminous Es,
the role of gas dissipation must be significantly reduced at high masses 
\citep{bender92,khochfar05,naab06d,kang07}, 
and recent ETG-ETG merger simulations
have demonstrated this numerically \citep{naab06a}. 
Figures \ref{fig:prog.type} and \ref{fig:cm.prog}
show that 90\% of the progenitors in this study
have concentrated light profiles and red-sequence colors, both common
attributes of ETGs, with little or no cold gas
content. In addition, the tidal signatures of the bulk of these massive
mergers (see Figs. \ref{fig:dpairs1} \& \ref{fig:dpairs2}) 
match those of observed \citep{bell06a} and simulated 
\citep{naab06a} major dissipationless (or gas-poor) merging of ETGs.
Thus, our sample represents a more than
order-of-magnitude increase in the number of such known systems with $z<0.2$,
and demonstrates that dissipationless merging is indeed an important channel for
the formation of massive galaxies.

Finally, we compare the observed high fraction of ETG-ETG mergers 
($f_{\rm ETG-ETG}=0.9$)
with several semi-analytic predictions.
Recall that we have looked for signs of interaction in $>200$
major pairs from a total sample of $>5000$ massive galaxies
(i.e., sampM), yet
only 10\% of the 38 mergers we identify could possibly form a
${\rm M}_{\rm star}>10^{11}{\rm M}_{\sun}$ remnant by other than an ETG-ETG
merger.
The progenitor morphologies of this study best match the predictions of
\citet{khochfar03}, who find $f_{\rm E-E}=0.75$ for the last major merger
of $4L^{\ast}$ remnants, independent of environment. We find much larger
ETG-ETG fractions than \citet{naab06a} who predict only 20-35\%
(also independent of environment)
over the estimated mass range of our merger remnants
($11.1<\log_{10}{({\rm M}_{\rm star}/{\rm M}_{\sun})}<11.7$), 
and \citet{kang07} who predict $f_{\rm ETG-ETG}<0.1$ for 
$\log_{10}{({\rm M}_{\rm star}/{\rm M}_{\sun})}>11$. We note that these
predicted progenitor morphologies for present-day Es are based on
the final major mergers that could occur over a large redshift range
out to $z\sim1$, which could be different in nature to those that occur
in the short time interval that we observe.
Moreover, we focus on high-density environments known to have very few
massive late-type (blue) galaxies \citep{butcher78a}, which might explain the
low number of
``mixed'' (early-late or elliptical-spiral) mergers that we find.
Hence, for these models to be consistent with our data,
either (1) the $f_{\rm ETG-ETG}$ of present-day major
mergers depends on halo mass (i.e., environment), 
or (2) the relative importance of
major mixed mergers has decreased significantly since $z=1$.

\subsection{Estimating Stellar Mass Accretion Rates}

The existence of massive dissipationless mergers at low redshift is direct
observational evidence that the growth of 
${\rm M}_{\rm star}>10^{11}{\rm M}_{\sun}$
galaxies continues at present times in agreement with many 
cosmologically-motivated simulations 
\citep{khochfar05,delucia06,kaviraj07,kang07}.
Moreover, even under conservative assumptions that limit the
amount of companion mass that is added to massive CEN galaxies, all
of our sample will still result in remnants with 
${\rm M}_{\rm star}>10^{11}{\rm M}_{\sun}$.
Previously, the observational evidence for recent merger-based assembly
of $z\sim0$ massive Es was limited to luminous/massive galaxy clustering 
statistics \citep{masjedi06,bell06b,masjedi07} or post-merger signatures that
cannot distinguish between minor and major merging; e.g., tidal
shells \citep{malin83}, fine structure \citep{schweizer92}, faint tidal
features \citep{vandokkum05,mihos05}, or
kinematic/photometric properties \citep[e.g.,][]{kang07}.
With the merger sample presented here we
can quantify directly the amount of growth, occurring in dense environments,
at the high-mass end of the stellar mass function. 

Going from the observed merger counts to an inferred merger rate
is limited mostly by the uncertainty in the merger timescale ($t_{\rm merg}$)
that one assumes.
Numerical models show that
the time interval for two galaxies to interact and finally merge
into a single remnant depends critically on the orbital parameters,
progenitor mass ratios and densities, and the degree to which the merger is
dissipationless. 
For major mergers of massive galaxies 
a number of different $t_{\rm merg}$ have been
put forth in the literature based on simple orbital timescale
arguments. For example, \citet{masjedi06} 
derived a reasonable lower limit of $t_{\rm merg}=0.2$ Gyr
for a close ($d_{1,2}=10$ kpc) pair of LRG galaxies with a velocity 
dispersion of $\sigma=200$ \kms.
Naturally, bound pairs with $d_{1,2}>10$ kpc separation
will take longer to merge. \citet{bell06b} made a similar calculation
for somewhat less-massive galaxies typically separated by
$d_{1,2}=15$ kpc and estimated $t_{\rm merg}=0.4$ Gyr and argued for at least
a factor of two uncertainty in this time. 
The mergers in this study have an average
projected separation of 15.5 kpc (see Fig. \ref{fig:pair.mergVnon}),
so in what follows, we adopt $t_{\rm merg}=0.4^{+0.4}_{-0.2}$ Gyr with 
conservative error bars that encompass the range of uncertainties
discussed in the literature.

Here, we compute the rate of stellar mass accretion by major merging
onto massive galaxies
in large groups. First, we find that the
total mass accreted onto the centers of the $N_{\rm CEN}=845$ halos that
we study is $\sum{ f{\rm M}_{{\rm s},i}}=3.9(3.5)\times10^{12} {\rm M}_{\sun}$,
if we include (exclude) the four SAT-SAT mergers at their host's dynamical 
center (see \S \ref{sec:cenmerging}). 
${\rm M}_{{\rm s},i}$ is the stellar mass of the
secondary (SAT) galaxy in the $i^{\rm th}$ CEN-SAT merger,
and $f$ is the fraction of ${\rm M}_{{\rm s},i}$ that winds up as part
of the CEN galaxy.
The rate of stellar mass buildup per massive CEN galaxy is therefore
\begin{equation}
\dot{{\rm M}}_{\rm CEN} = \frac{\sum{ f{\rm M}_{{\rm s},i} }}{N_{\rm CEN}} \times \frac{1}{t_{\rm merg}} ,
\label{eq:4}
\end{equation}
or between $1.0^{+1.0}_{-0.5}\times10^{10}{\rm M}_{\sun}{\rm Gyr}^{-1}$
and $1.2^{+1.1}_{-0.6}\times10^{10}{\rm M}_{\sun}{\rm Gyr}^{-1}$,
depending on which sample of CEN-SAT mergers that we consider. 
The lopsided error bars result from
the range of accretion rates for $t_{\rm merg}=0.4^{+0.4}_{-0.2}$ Gyr,
as described above.
If we divide all of these accretion rates by $2.69\times10^{11} {\rm M}_{\sun}$,
the average stellar mass of the 845 CEN galaxies in this
study, we find that each CEN
is growing by 1--9\% per Gyr.
Finally, these values can be decreased by assuming $f<1$ in (\ref{eq:4}),
but as we discuss in \S \ref{sec:cmplane}, 
$f=0.5$ represents a likely lower limit.

Likewise, the total stellar mass accreted onto all galaxies in sampM is
$\sum{ f{\rm M}_{{\rm s},i}}+\sum{ {\rm M}_{{\rm s},j}}=5.1\times10^{12} {\rm M}_{\sun}$,
where ${\rm M}_{{\rm s},j}$ is the mass of the
secondary (SAT) galaxy in the $j^{\rm th}$ SAT-SAT merger.
Therefore, the growth per ${\rm M}_{\rm star}\geq5\times10^{10} {\rm M}_{\sun}$
galaxy in high-mass groups is
\begin{equation}
\dot{{\rm M}}_{(\geq5\times10^{10} {\rm M}_{\sun})} = \frac{\sum{ f{\rm M}_{{\rm s},j} } + \sum{ {\rm M}_{{\rm s},j}} }{ (N_{\rm CEN}+N_{\rm SAT}-N_{\rm s,sampM})} \times \frac{1}{t_{\rm merg}} ,
\end{equation}
where $N_{\rm s,sampM}=12$ is the number of secondary SAT galaxies in sampM
that are involved in major mergers and must be subtracted to avoid 
double counting. We find
$\dot{{\rm M}}_{(\geq5\times10^{10} {\rm M}_{\sun})}=2.4^{+2.4}_{-1.2}\times10^{9}{\rm M}_{\sun}{\rm Gyr}^{-1}$; if we assume $f=0.5$ for CEN-SAT mergers only we find
$\dot{{\rm M}}_{(\geq5\times10^{10} {\rm M}_{\sun})}=1.6^{+1.5}_{-0.7}\times10^{9}{\rm M}_{\sun}{\rm Gyr}^{-1}$.
Given that the average stellar mass of sampM galaxies is
$1.04\times10^{11} {\rm M}_{\sun}$, we find that every massive
galaxy is growing by 1--5\% per Gyr.
Even though SAT-SAT mergers may occur as frequently as
CEN-SAT mergers in these massive groups, the centers are where
much of the mass growth takes place. It is clear from Figure \ref{fig:prog.mass}
that mostly only ${\rm M}_{\rm star}>10^{11} {\rm M}_{\sun}$
galaxies build up in mass by major mergers in groups 
with ${\rm M}_{\rm halo}>2.5\times10^{13}{\rm M}_{\sun}$.
In contrast, we find few mergers among the 
$5\times 10^{10}<{\rm M}_{\rm star}<10^{11} {\rm M}_{\sun}$ galaxies
in these high-mass groups, which make up the bulk (60\%) of sampM.
This suggests
that if major merging is playing an important role in the strong mass growth 
observed on the red sequence below M*
\citep{bell04b,blanton06,borch06,faber07,brown07},
it is occurring in lower-mass groups than we study here.


Rather than mass growth rates we can use the same line of reasoning
to estimate massive galaxy-galaxy merging rates of 
$(21+4)/845/t_{\rm merg}=0.074^{+0.074}_{-0.037}{\rm Gyr}^{-1}$ for CEN-SAT and
$(38+7)/(845+4531-12)/t_{\rm merg}=0.021^{+0.021}_{-0.011}{\rm Gyr}^{-1}$ for 
all galaxies in sampM.
For these estimates we included the seven additional major mergers
(4 CEN, 3 SAT) we identified by their highly-disturbed appearance.
\citet{masjedi06} found a strict upper limit to the LRG-LRG rate of only
$0.006{\rm Gyr}^{-1}$. 
We estimate that LRGs have a stellar mass range of
$11.4<\log_{10}{({\rm M}_{\rm star}/{\rm M}_{\sun})}<12.0$, based on
typical red-sequence colors and luminosities between $4L^{\ast}$ and 
$25L^{\ast}$. Within these mass limits, we find a merger rate of
$5/462/t_{\rm merg}=0.027^{+0.027}_{-0.014}{\rm Gyr}^{-1}$ on the
red sequence, or
2--9 times the LRG-LRG rate.
In Table \ref{mstar_counts},
we show that the high-mass groups that we study
contain $>70\%$ of the very-massive, red galaxy population in the
$z\leq0.12$ volume of DR2, with the vast majority being CENs. Yet, the same
group selection contains only 30\% of the population of
$11.4<\log_{10}{({\rm M}_{\rm star}/{\rm M}_{\sun})}<11.6$ systems.
These numbers show that a significant portion of the local counterparts 
to LRGs are found in groups with 
${\rm M}_{\rm halo}<2.5\times10^{13}{\rm M}_{\sun}$.
Therefore, we conclude that LRG-LRG merging occurs more
frequently in the more massive groups.

\section{Summary}
Using the SDSS DR2 group catalog we probe a sufficiently large enough volume
of the low-redshift universe to identify major mergers that will produce
${\rm M}_{\rm star}\geq10^{11} {\rm M}_{\sun}$ galaxies in large
groups and clusters. We find 45 massive mergers in a complete sample 
of more than 5000 galaxies
with ${\rm M}_{\rm star}\geq5\times10^{10} {\rm M}_{\sun}$ that reside
in 845 groups with ${\rm M}_{\rm halo}>2.5\times10^{13}{\rm M}_{\sun}$.
We identify 38 pairs of merging galaxies such that both systems
exhibit asymmetric features consistent with mutual tidal interactions,
and another seven mergers that have disturbed morphologies and
semi-resolved double nuclei. This work provides
the first direct evidence for present-day massive mergers, and complements
existing studies at higher redshifts \citep{vandokkum99,bell06a,lotz06,rines07}.
With this sample,
we provide new empirical constraints on the progenitor nature,
the environmental dependence, and the
stellar mass growth rate of merger-driven assembly of high-mass 
galaxies. We summarize our results as follows:
\begin{itemize}
\item
Mergers, as defined here, make up only 16\% of the
major pairs of massive galaxies with a maximum projected 
separation of 30 kpc.
\item
An important percentage (70\%) of these mergers would be lost in
an automated search of spec-spec pairs as a result of the known
spectroscopic incompleteness of the SDSS in dense environments.
\item
90\% of the mergers are between two red-sequence galaxies with
concentrated (spheroid-dominated) morphologies, and
broad tidal asymmetries like those seen in observations and in
simulations of major dissipationless merging of spheroidal galaxies
\citep{naab06a,bell06a}.
\item
Two thirds of the mergers have progenitor mass ratios of 1:1 to 2:1, despite
a complete search of major pairs down to 4:1, indicating that near
equal-mass merging is preferred in high-density environments.
\item
Mergers at the centers of massive groups are more common than between
two SAT galaxies, but the latter are also identified and are
morphologically indistinguishable from CEN-SAT mergers.
We argue that SAT-SAT mergers could identify the dynamical
centers of large subhalos that have recently been accreted by their 
host halo, rather than the centers of distinct halos seen in projection.
\item
The frequency that massive SATs have a major merger with a
more-massive CEN or SAT galaxy
decreases with halo mass in a manner that is qualitatively consistent
with the expectations of dynamical friction.
\item
Based on reasonable assumptions, the centers of massive halos in the
present-day universe are growing in stellar mass by 1--9\% per Gyr on average,
through major mergers as we observe here.
\item
Red galaxies with ${\rm M}_{\rm star}\geq2.5\times10^{11} {\rm M}_{\sun}$, 
which are comparable to LRGs, merge with their counterparts
in these high-mass groups at a rate that is 2--9 times 
higher than that found for all LRG-LRG merging by
\citet{masjedi06}.
\end{itemize}
It is becoming clear that gas-poor, major merging between massive red and
bulge-dominated galaxies is an important mechanism for producing the
most-massive galaxies. Using the SDSS we have demonstrated that the
centers of dark-matter halos are the preferred environment for building
these giants. Moreover, this analysis shows that our
technique for identifying such mergers is very promising for future
studies of much larger samples.

\acknowledgements  
We made extensive use of the SDSS SkyServer Tools
({\texttt http://cas.sdss.org/astro/en/tools/}).
Thanks to discussions with Chien Peng, Martin Weinberg, Eric Bell, and Yu Lu.
D.\ H.\ M.\ and N.\ K.\
acknowledge support from the National Aeronautics
and Space Administration (NASA) under LTSA Grant NAG5-13102
issued through the Office of Space Science.
Funding for the SDSS
has been provided by the Alfred P.\ Sloan Foundation, the
Participating Institutions, the National Aeronautics and Space Administration,
the National Science Foundation, the U.S. Department of Energy,
the Japanese Monbukagakusho, and the Max Planck Society.  The SDSS
Web site is {\texttt http://www.sdss.org/}.  The SDSS is managed by the
Astrophysical Research Consortium (ARC) for the Participating Institutions,
which are The University of Chicago, Fermilab, the Institute
for Advanced Study, the Japan Participation Group, The Johns Hopkins
University, Los Alamos National Laboratory,
the Max-Planck-Institute for Astronomy (MPIA), the Max-Planck-Institute for
Astrophysics (MPA), New Mexico State University, University of Pittsburgh,
Princeton University, the United States Naval Observatory, and
the University of Washington.
This publication also made use of NASA's Astrophysics Data System
Bibliographic Services.

\bibliographystyle{/home/dmac/Papers/apj}
\bibliography{/home/dmac/Papers/references,/home/dmac/Papers/preprints}

\begin{thebibliography}{109}
\expandafter\ifx\csname natexlab\endcsname\relax\def\natexlab#1{#1}\fi

\bibitem[{{Abazajian} {et~al.}(2004){Abazajian}, {Adelman-McCarthy}, {Ag{\"
  u}eros}, {Allam}, {Anderson}, {Anderson}, {Annis}, {Bahcall}, {Baldry},
  {Bastian}, \& {et al.}}]{abazajian04}
{Abazajian}, K., {Adelman-McCarthy}, J.~K., {Ag{\" u}eros}, M.~A., {Allam},
  S.~S., {Anderson}, K.~S.~J., {Anderson}, S.~F., {Annis}, J., {Bahcall},
  N.~A., {Baldry}, I.~K., {Bastian}, S., \& {et al.} 2004, \aj, 128, 502

\bibitem[{{Aragon-Salamanca} {et~al.}(1998){Aragon-Salamanca}, {Baugh}, \&
  {Kauffmann}}]{aragon98}
{Aragon-Salamanca}, A., {Baugh}, C.~M., \& {Kauffmann}, G. 1998, \mnras, 297,
  427

\bibitem[{{Barnes}(1988)}]{barnes88}
{Barnes}, J.~E. 1988, \apj, 331, 699

\bibitem[{{Barnes} \& {Hernquist}(1992)}]{barnes92b}
{Barnes}, J.~E., \& {Hernquist}, L. 1992, \araa, 30, 705

\bibitem[{{Barnes} \& {Hernquist}(1996)}]{barnes96}
---. 1996, \apj, 471, 115

\bibitem[{{Bell} {et~al.}(2004{\natexlab{a}}){Bell}, {McIntosh}, {Barden},
  {Wolf}, {Caldwell}, {Rix}, {Beckwith}, {Borch}, {H{\" a}ussler}, {Jahnke},
  {Jogee}, {Meisenheimer}, {Peng}, {Sanchez}, {Somerville}, \&
  {Wisotzki}}]{bell04a}
{Bell}, E.~F., {McIntosh}, D.~H., {Barden}, M., {Wolf}, C., {Caldwell},
  J.~A.~R., {Rix}, H.-W., {Beckwith}, S.~V.~W., {Borch}, A., {H{\" a}ussler},
  B., {Jahnke}, K., {Jogee}, S., {Meisenheimer}, K., {Peng}, C., {Sanchez},
  S.~F., {Somerville}, R.~S., \& {Wisotzki}, L. 2004{\natexlab{a}}, \apjl, 600,
  L11

\bibitem[{{Bell} {et~al.}(2003){Bell}, {McIntosh}, {Katz}, \&
  {Weinberg}}]{bell03b}
{Bell}, E.~F., {McIntosh}, D.~H., {Katz}, N., \& {Weinberg}, M.~D. 2003, \apjs,
  149, 289

\bibitem[{{Bell} {et~al.}(2006{\natexlab{a}}){Bell}, {Naab}, {McIntosh},
  {Somerville}, {Caldwell}, {Barden}, {Wolf}, {Rix}, {Beckwith}, {Borch},
  {H{\"a}ussler}, {Heymans}, {Jahnke}, {Jogee}, {Koposov}, {Meisenheimer},
  {Peng}, {Sanchez}, \& {Wisotzki}}]{bell06a}
{Bell}, E.~F., {Naab}, T., {McIntosh}, D.~H., {Somerville}, R.~S., {Caldwell},
  J.~A.~R., {Barden}, M., {Wolf}, C., {Rix}, H.-W., {Beckwith}, S.~V., {Borch},
  A., {H{\"a}ussler}, B., {Heymans}, C., {Jahnke}, K., {Jogee}, S., {Koposov},
  S., {Meisenheimer}, K., {Peng}, C.~Y., {Sanchez}, S.~F., \& {Wisotzki}, L.
  2006{\natexlab{a}}, \apj, 640, 241

\bibitem[{{Bell} {et~al.}(2005){Bell}, {Papovich}, {Wolf}, {Le Floc'h},
  {Caldwell}, {Barden}, {Egami}, {McIntosh}, {Meisenheimer},
  {P{\'e}rez-Gonz{\'a}lez}, {Rieke}, {Rieke}, {Rigby}, \& {Rix}}]{bell05a}
{Bell}, E.~F., {Papovich}, C., {Wolf}, C., {Le Floc'h}, E., {Caldwell},
  J.~A.~R., {Barden}, M., {Egami}, E., {McIntosh}, D.~H., {Meisenheimer}, K.,
  {P{\'e}rez-Gonz{\'a}lez}, P.~G., {Rieke}, G.~H., {Rieke}, M.~J., {Rigby},
  J.~R., \& {Rix}, H.-W. 2005, \apj, 625, 23

\bibitem[{{Bell} {et~al.}(2006{\natexlab{b}}){Bell}, {Phleps}, {Somerville},
  {Wolf}, {Borch}, \& {Meisenheimer}}]{bell06b}
{Bell}, E.~F., {Phleps}, S., {Somerville}, R.~S., {Wolf}, C., {Borch}, A., \&
  {Meisenheimer}, K. 2006{\natexlab{b}}, \apj, 652, 270

\bibitem[{{Bell} {et~al.}(2004{\natexlab{b}}){Bell}, {Wolf}, {Meisenheimer},
  {Rix}, {Borch}, {Dye}, {Kleinheinrich}, {Wisotzki}, \& {McIntosh}}]{bell04b}
{Bell}, E.~F., {Wolf}, C., {Meisenheimer}, K., {Rix}, H.-W., {Borch}, A.,
  {Dye}, S., {Kleinheinrich}, M., {Wisotzki}, L., \& {McIntosh}, D.~H.
  2004{\natexlab{b}}, \apj, 608, 752

\bibitem[{{Bender}(1988)}]{bender88}
{Bender}, R. 1988, \aap, 193, L7

\bibitem[{{Bender} {et~al.}(1992){Bender}, {Burstein}, \& {Faber}}]{bender92}
{Bender}, R., {Burstein}, D., \& {Faber}, S.~M. 1992, \apj, 399, 462

\bibitem[{{Berrier} {et~al.}(2006){Berrier}, {Bullock}, {Barton}, {Guenther},
  {Zentner}, \& {Wechsler}}]{berrier06}
{Berrier}, J.~C., {Bullock}, J.~S., {Barton}, E.~J., {Guenther}, H.~D.,
  {Zentner}, A.~R., \& {Wechsler}, R.~H. 2006, \apj, 652, 56

\bibitem[{{Blanton}(2006)}]{blanton06}
{Blanton}, M.~R. 2006, \apj, 648, 268

\bibitem[{{Blanton} {et~al.}(2003{\natexlab{a}}){Blanton}, {Brinkmann},
  {Csabai}, {Doi}, {Eisenstein}, {Fukugita}, {Gunn}, {Hogg}, \&
  {Schlegel}}]{blanton03b}
{Blanton}, M.~R., {Brinkmann}, J., {Csabai}, I., {Doi}, M., {Eisenstein}, D.,
  {Fukugita}, M., {Gunn}, J.~E., {Hogg}, D.~W., \& {Schlegel}, D.~J.
  2003{\natexlab{a}}, \aj, 125, 2348

\bibitem[{{Blanton} {et~al.}(2003{\natexlab{b}}){Blanton}, {Hogg}, {Bahcall},
  {Baldry}, {Brinkmann}, {Csabai}, {Eisenstein}, {Fukugita}, {Gunn}, {Ivezi{\'
  c}}, {Lamb}, {Lupton}, {Loveday}, {Munn}, {Nichol}, {Okamura}, {Schlegel},
  {Shimasaku}, {Strauss}, {Vogeley}, \& {Weinberg}}]{blanton03d}
{Blanton}, M.~R., {Hogg}, D.~W., {Bahcall}, N.~A., {Baldry}, I.~K.,
  {Brinkmann}, J., {Csabai}, I., {Eisenstein}, D., {Fukugita}, M., {Gunn},
  J.~E., {Ivezi{\' c}}, {\v Z}., {Lamb}, D.~Q., {Lupton}, R.~H., {Loveday}, J.,
  {Munn}, J.~A., {Nichol}, R.~C., {Okamura}, S., {Schlegel}, D.~J.,
  {Shimasaku}, K., {Strauss}, M.~A., {Vogeley}, M.~S., \& {Weinberg}, D.~H.
  2003{\natexlab{b}}, \apj, 594, 186

\bibitem[{{Blanton} {et~al.}(2003{\natexlab{c}}){Blanton}, {Hogg}, {Bahcall},
  {Brinkmann}, {Britton}, {Connolly}, {Csabai}, {Fukugita}, {Loveday},
  {Meiksin}, {Munn}, {Nichol}, {Okamura}, {Quinn}, {Schneider}, {Shimasaku},
  {Strauss}, {Tegmark}, {Vogeley}, \& {Weinberg}}]{blanton03c}
{Blanton}, M.~R., {Hogg}, D.~W., {Bahcall}, N.~A., {Brinkmann}, J., {Britton},
  M., {Connolly}, A.~J., {Csabai}, I., {Fukugita}, M., {Loveday}, J.,
  {Meiksin}, A., {Munn}, J.~A., {Nichol}, R.~C., {Okamura}, S., {Quinn}, T.,
  {Schneider}, D.~P., {Shimasaku}, K., {Strauss}, M.~A., {Tegmark}, M.,
  {Vogeley}, M.~S., \& {Weinberg}, D.~H. 2003{\natexlab{c}}, \apj, 592, 819

\bibitem[{{Blanton} {et~al.}(2005){Blanton}, {Schlegel}, {Strauss},
  {Brinkmann}, {Finkbeiner}, {Fukugita}, {Gunn}, {Hogg}, {Ivezi{\'c}}, {Knapp},
  {Lupton}, {Munn}, {Schneider}, {Tegmark}, \& {Zehavi}}]{blanton05}
{Blanton}, M.~R., {Schlegel}, D.~J., {Strauss}, M.~A., {Brinkmann}, J.,
  {Finkbeiner}, D., {Fukugita}, M., {Gunn}, J.~E., {Hogg}, D.~W., {Ivezi{\'c}},
  {\v Z}., {Knapp}, G.~R., {Lupton}, R.~H., {Munn}, J.~A., {Schneider}, D.~P.,
  {Tegmark}, M., \& {Zehavi}, I. 2005, \aj, 129, 2562

\bibitem[{{Blumenthal} {et~al.}(1984){Blumenthal}, {Faber}, {Primack}, \&
  {Rees}}]{blumenthal84}
{Blumenthal}, G.~R., {Faber}, S.~M., {Primack}, J.~R., \& {Rees}, M.~J. 1984,
  \nat, 311, 517

\bibitem[{{Borch} {et~al.}(2006){Borch}, {Meisenheimer}, {Bell}, {Rix}, {Wolf},
  {Dye}, {Kleinheinrich}, {Kovacs}, \& {Wisotzki}}]{borch06}
{Borch}, A., {Meisenheimer}, K., {Bell}, E.~F., {Rix}, H.-W., {Wolf}, C.,
  {Dye}, S., {Kleinheinrich}, M., {Kovacs}, Z., \& {Wisotzki}, L. 2006, \aap,
  453, 869

\bibitem[{{Boylan-Kolchin} {et~al.}(2006){Boylan-Kolchin}, {Ma}, \&
  {Quataert}}]{boylan06}
{Boylan-Kolchin}, M., {Ma}, C.-P., \& {Quataert}, E. 2006, \mnras, 369, 1081

\bibitem[{{Brown} {et~al.}(2007){Brown}, {Dey}, {Jannuzi}, {Brand}, {Benson},
  {Brodwin}, {Croton}, \& {Eisenhardt}}]{brown07}
{Brown}, M.~J.~I., {Dey}, A., {Jannuzi}, B.~T., {Brand}, K., {Benson}, A.~J.,
  {Brodwin}, M., {Croton}, D.~J., \& {Eisenhardt}, P.~R. 2007, \apj, 654, 858

\bibitem[{{Bundy} {et~al.}(2004){Bundy}, {Fukugita}, {Ellis}, {Kodama}, \&
  {Conselice}}]{bundy04}
{Bundy}, K., {Fukugita}, M., {Ellis}, R.~S., {Kodama}, T., \& {Conselice},
  C.~J. 2004, \apjl, 601, L123

\bibitem[{{Burstein} {et~al.}(1997){Burstein}, {Bender}, {Faber}, \&
  {Nolthenius}}]{burstein97}
{Burstein}, D., {Bender}, R., {Faber}, S., \& {Nolthenius}, R. 1997, \aj, 114,
  1365

\bibitem[{{Butcher} \& {Oemler}(1978)}]{butcher78a}
{Butcher}, H., \& {Oemler}, A. 1978, \apj, 219, 18

\bibitem[{{Carlberg} {et~al.}(2000){Carlberg}, {Cohen}, {Patton}, {Blandford},
  {Hogg}, {Yee}, {Morris}, {Lin}, {Hall}, {Sawicki}, {Wirth}, {Cowie}, {Hu}, \&
  {Songaila}}]{carlberg00}
{Carlberg}, R.~G., {Cohen}, J.~G., {Patton}, D.~R., {Blandford}, R., {Hogg},
  D.~W., {Yee}, H.~K.~C., {Morris}, S.~L., {Lin}, H., {Hall}, P.~B., {Sawicki},
  M., {Wirth}, G.~D., {Cowie}, L.~L., {Hu}, E., \& {Songaila}, A. 2000, \apjl,
  532, L1

\bibitem[{{Carlberg} {et~al.}(1994){Carlberg}, {Pritchet}, \&
  {Infante}}]{carlberg94}
{Carlberg}, R.~G., {Pritchet}, C.~J., \& {Infante}, L. 1994, \apj, 435, 540

\bibitem[{{Cavaliere} {et~al.}(1992){Cavaliere}, {Colafrancesco}, \&
  {Menci}}]{cavaliere92}
{Cavaliere}, A., {Colafrancesco}, S., \& {Menci}, N. 1992, \apj, 392, 41

\bibitem[{{Cole} {et~al.}(2000){Cole}, {Lacey}, {Baugh}, \& {Frenk}}]{cole00}
{Cole}, S., {Lacey}, C.~G., {Baugh}, C.~M., \& {Frenk}, C.~S. 2000, \mnras,
  319, 168

\bibitem[{Conroy {et~al.}(2007)Conroy, Wechsler, \& Kravtsov}]{conroy07c}
Conroy, C., Wechsler, R.~H., \& Kravtsov, A.~V. 2007

\bibitem[{{Conselice} {et~al.}(2003){Conselice}, {Bershady}, {Dickinson}, \&
  {Papovich}}]{conselice03}
{Conselice}, C.~J., {Bershady}, M.~A., {Dickinson}, M., \& {Papovich}, C. 2003,
  \aj, 126, 1183

\bibitem[{{Cooray} \& {Milosavljevi{\'c}}(2005)}]{cooray05d}
{Cooray}, A., \& {Milosavljevi{\'c}}, M. 2005, \apjl, 627, L85

\bibitem[{{Cox} {et~al.}(2006){Cox}, {Dutta}, {Di Matteo}, {Hernquist},
  {Hopkins}, {Robertson}, \& {Springel}}]{cox06}
{Cox}, T.~J., {Dutta}, S.~N., {Di Matteo}, T., {Hernquist}, L., {Hopkins},
  P.~F., {Robertson}, B., \& {Springel}, V. 2006, \apj, 650, 791

\bibitem[{{Davies} {et~al.}(1983){Davies}, {Efstathiou}, {Fall}, {Illingworth},
  \& {Schechter}}]{davies83}
{Davies}, R.~L., {Efstathiou}, G., {Fall}, S.~M., {Illingworth}, G., \&
  {Schechter}, P.~L. 1983, \apj, 266, 41

\bibitem[{{Davis} {et~al.}(1985){Davis}, {Efstathiou}, {Frenk}, \&
  {White}}]{davis85}
{Davis}, M., {Efstathiou}, G., {Frenk}, C.~S., \& {White}, S.~D.~M. 1985, \apj,
  292, 371

\bibitem[{{De Lucia} \& {Blaizot}(2007)}]{delucia07}
{De Lucia}, G., \& {Blaizot}, J. 2007, \mnras, 375, 2

\bibitem[{{De Lucia} {et~al.}(2006){De Lucia}, {Springel}, {White}, {Croton},
  \& {Kauffmann}}]{delucia06}
{De Lucia}, G., {Springel}, V., {White}, S.~D.~M., {Croton}, D., \&
  {Kauffmann}, G. 2006, \mnras, 366, 499

\bibitem[{{Dressler}(1980)}]{dressler80a}
{Dressler}, A. 1980, \apj, 236, 351

\bibitem[{{Dubinski}(1998)}]{dubinski98}
{Dubinski}, J. 1998, \apj, 502, 141

\bibitem[{{Dubinski} {et~al.}(1996){Dubinski}, {Mihos}, \&
  {Hernquist}}]{dubinski96}
{Dubinski}, J., {Mihos}, J.~C., \& {Hernquist}, L. 1996, \apj, 462, 576

\bibitem[{{Eisenstein} {et~al.}(2001){Eisenstein}, {Annis}, {Gunn}, {Szalay},
  {Connolly}, {Nichol}, {Bahcall}, {Bernardi}, {Burles}, {Castander},
  {Fukugita}, {Hogg}, {Ivezi{\'c}}, {Knapp}, {Lupton}, {Narayanan}, {Postman},
  {Reichart}, {Richmond}, {Schneider}, {Schlegel}, {Strauss}, {SubbaRao},
  {Tucker}, {Vanden Berk}, {Vogeley}, {Weinberg}, \& {Yanny}}]{eisenstein01}
{Eisenstein}, D.~J., {Annis}, J., {Gunn}, J.~E., {Szalay}, A.~S., {Connolly},
  A.~J., {Nichol}, R.~C., {Bahcall}, N.~A., {Bernardi}, M., {Burles}, S.,
  {Castander}, F.~J., {Fukugita}, M., {Hogg}, D.~W., {Ivezi{\'c}}, {\v Z}.,
  {Knapp}, G.~R., {Lupton}, R.~H., {Narayanan}, V., {Postman}, M., {Reichart},
  D.~E., {Richmond}, M., {Schneider}, D.~P., {Schlegel}, D.~J., {Strauss},
  M.~A., {SubbaRao}, M., {Tucker}, D.~L., {Vanden Berk}, D., {Vogeley}, M.~S.,
  {Weinberg}, D.~H., \& {Yanny}, B. 2001, \aj, 122, 2267

\bibitem[{{Faber} {et~al.}(1997){Faber}, {Tremaine}, {Ajhar}, {Byun},
  {Dressler}, {Gebhardt}, {Grillmair}, {Kormendy}, {Lauer}, \&
  {Richstone}}]{faber97}
{Faber}, S.~M., {Tremaine}, S., {Ajhar}, E.~A., {Byun}, Y., {Dressler}, A.,
  {Gebhardt}, K., {Grillmair}, C., {Kormendy}, J., {Lauer}, T.~R., \&
  {Richstone}, D. 1997, \aj, 114, 1771

\bibitem[{{Faber} {et~al.}(2007){Faber}, {Willmer}, {Wolf}, {Koo}, {Weiner},
  {Newman}, {Im}, {Coil}, {Conroy}, {Cooper}, {Davis}, {Finkbeiner}, {Gerke},
  {Gebhardt}, {Groth}, {Guhathakurta}, {Harker}, {Kaiser}, {Kassin},
  {Kleinheinrich}, {Konidaris}, {Kron}, {Lin}, {Luppino}, {Madgwick},
  {Meisenheimer}, {Noeske}, {Phillips}, {Sarajedini}, {Schiavon}, {Simard},
  {Szalay}, {Vogt}, \& {Yan}}]{faber07}
{Faber}, S.~M., {Willmer}, C.~N.~A., {Wolf}, C., {Koo}, D.~C., {Weiner}, B.~J.,
  {Newman}, J.~A., {Im}, M., {Coil}, A.~L., {Conroy}, C., {Cooper}, M.~C.,
  {Davis}, M., {Finkbeiner}, D.~P., {Gerke}, B.~F., {Gebhardt}, K., {Groth},
  E.~J., {Guhathakurta}, P., {Harker}, J., {Kaiser}, N., {Kassin}, S.,
  {Kleinheinrich}, M., {Konidaris}, N.~P., {Kron}, R.~G., {Lin}, L., {Luppino},
  G., {Madgwick}, D.~S., {Meisenheimer}, K., {Noeske}, K.~G., {Phillips},
  A.~C., {Sarajedini}, V.~L., {Schiavon}, R.~P., {Simard}, L., {Szalay}, A.~S.,
  {Vogt}, N.~P., \& {Yan}, R. 2007, \apj, 665, 265

\bibitem[{{Fukugita} {et~al.}(1996){Fukugita}, {Ichikawa}, {Gunn}, {Doi},
  {Shimasaku}, \& {Schneider}}]{fukugita96}
{Fukugita}, M., {Ichikawa}, T., {Gunn}, J.~E., {Doi}, M., {Shimasaku}, K., \&
  {Schneider}, D.~P. 1996, \aj, 111, 1748

\bibitem[{{Gunn} {et~al.}(1998){Gunn}, {Carr}, {Rockosi}, {Sekiguchi}, {Berry},
  {Elms}, {de Haas}, {Ivezi{\'c}}, {Knapp}, {Lupton}, {Pauls}, {Simcoe},
  {Hirsch}, {Sanford}, {Wang}, {York}, {Harris}, {Annis}, {Bartozek},
  {Boroski}, {Bakken}, {Haldeman}, {Kent}, {Holm}, {Holmgren}, {Petravick},
  {Prosapio}, {Rechenmacher}, {Doi}, {Fukugita}, {Shimasaku}, {Okada}, {Hull},
  {Siegmund}, {Mannery}, {Blouke}, {Heidtman}, {Schneider}, {Lucinio}, \&
  {Brinkman}}]{gunn98}
{Gunn}, J.~E., {Carr}, M., {Rockosi}, C., {Sekiguchi}, M., {Berry}, K., {Elms},
  B., {de Haas}, E., {Ivezi{\'c}}, {\v Z}., {Knapp}, G., {Lupton}, R., {Pauls},
  G., {Simcoe}, R., {Hirsch}, R., {Sanford}, D., {Wang}, S., {York}, D.,
  {Harris}, F., {Annis}, J., {Bartozek}, L., {Boroski}, W., {Bakken}, J.,
  {Haldeman}, M., {Kent}, S., {Holm}, S., {Holmgren}, D., {Petravick}, D.,
  {Prosapio}, A., {Rechenmacher}, R., {Doi}, M., {Fukugita}, M., {Shimasaku},
  K., {Okada}, N., {Hull}, C., {Siegmund}, W., {Mannery}, E., {Blouke}, M.,
  {Heidtman}, D., {Schneider}, D., {Lucinio}, R., \& {Brinkman}, J. 1998, \aj,
  116, 3040

\bibitem[{{Gunn} {et~al.}(2006){Gunn}, {Siegmund}, {Mannery}, {Owen}, {Hull},
  {Leger}, {Carey}, {Knapp}, {York}, {Boroski}, {Kent}, {Lupton}, {Rockosi},
  {Evans}, {Waddell}, {Anderson}, {Annis}, {Barentine}, {Bartoszek}, {Bastian},
  {Bracker}, {Brewington}, {Briegel}, {Brinkmann}, {Brown}, {Carr},
  {Czarapata}, {Drennan}, {Dombeck}, {Federwitz}, {Gillespie}, {Gonzales},
  {Hansen}, {Harvanek}, {Hayes}, {Jordan}, {Kinney}, {Klaene}, {Kleinman},
  {Kron}, {Kresinski}, {Lee}, {Limmongkol}, {Lindenmeyer}, {Long}, {Loomis},
  {McGehee}, {Mantsch}, {Neilsen}, {Neswold}, {Newman}, {Nitta}, {Peoples},
  {Pier}, {Prieto}, {Prosapio}, {Rivetta}, {Schneider}, {Snedden}, \&
  {Wang}}]{gunn06}
{Gunn}, J.~E., {Siegmund}, W.~A., {Mannery}, E.~J., {Owen}, R.~E., {Hull},
  C.~L., {Leger}, R.~F., {Carey}, L.~N., {Knapp}, G.~R., {York}, D.~G.,
  {Boroski}, W.~N., {Kent}, S.~M., {Lupton}, R.~H., {Rockosi}, C.~M., {Evans},
  M.~L., {Waddell}, P., {Anderson}, J.~E., {Annis}, J., {Barentine}, J.~C.,
  {Bartoszek}, L.~M., {Bastian}, S., {Bracker}, S.~B., {Brewington}, H.~J.,
  {Briegel}, C.~I., {Brinkmann}, J., {Brown}, Y.~J., {Carr}, M.~A.,
  {Czarapata}, P.~C., {Drennan}, C.~C., {Dombeck}, T., {Federwitz}, G.~R.,
  {Gillespie}, B.~A., {Gonzales}, C., {Hansen}, S.~U., {Harvanek}, M., {Hayes},
  J., {Jordan}, W., {Kinney}, E., {Klaene}, M., {Kleinman}, S.~J., {Kron},
  R.~G., {Kresinski}, J., {Lee}, G., {Limmongkol}, S., {Lindenmeyer}, C.~W.,
  {Long}, D.~C., {Loomis}, C.~L., {McGehee}, P.~M., {Mantsch}, P.~M.,
  {Neilsen}, Jr., E.~H., {Neswold}, R.~M., {Newman}, P.~R., {Nitta}, A.,
  {Peoples}, J.~J., {Pier}, J.~R., {Prieto}, P.~S., {Prosapio}, A., {Rivetta},
  C., {Schneider}, D.~P., {Snedden}, S., \& {Wang}, S.-i. 2006, \aj, 131, 2332

\bibitem[{{Hashimoto} \& {Oemler}(1999)}]{hashimoto99}
{Hashimoto}, Y., \& {Oemler}, A.~J. 1999, \apj, 510, 609

\bibitem[{{Hogg} {et~al.}(2002){Hogg}, {Blanton}, {Strateva}, {Bahcall},
  {Brinkmann}, {Csabai}, {Doi}, {Fukugita}, {Hennessy}, {Ivezi{\' c}}, {Knapp},
  {Lamb}, {Lupton}, {Munn}, {Nichol}, {Schlegel}, {Schneider}, \&
  {York}}]{hogg02}
{Hogg}, D.~W., {Blanton}, M., {Strateva}, I., {Bahcall}, N.~A., {Brinkmann},
  J., {Csabai}, I., {Doi}, M., {Fukugita}, M., {Hennessy}, G., {Ivezi{\' c}},
  {\v Z}., {Knapp}, G.~R., {Lamb}, D.~Q., {Lupton}, R., {Munn}, J.~A.,
  {Nichol}, R., {Schlegel}, D.~J., {Schneider}, D.~P., \& {York}, D.~G. 2002,
  \aj, 124, 646

\bibitem[{{Hogg} {et~al.}(2004){Hogg}, {Blanton}, {Brinchmann}, {Eisenstein},
  {Schlegel}, {Gunn}, {McKay}, {Rix}, {Bahcall}, {Brinkmann}, \&
  {Meiksin}}]{hogg04}
{Hogg}, D.~W., {Blanton}, M.~R., {Brinchmann}, J., {Eisenstein}, D.~J.,
  {Schlegel}, D.~J., {Gunn}, J.~E., {McKay}, T.~A., {Rix}, H.-W., {Bahcall},
  N.~A., {Brinkmann}, J., \& {Meiksin}, A. 2004, \apjl, 601, L29

\bibitem[{{Hogg} {et~al.}(2001){Hogg}, {Finkbeiner}, {Schlegel}, \&
  {Gunn}}]{hogg01}
{Hogg}, D.~W., {Finkbeiner}, D.~P., {Schlegel}, D.~J., \& {Gunn}, J.~E. 2001,
  \aj, 122, 2129

\bibitem[{{Hopkins} {et~al.}(2006){Hopkins}, {Hernquist}, {Cox}, {Di Matteo},
  {Robertson}, \& {Springel}}]{hopkins06b}
{Hopkins}, P.~F., {Hernquist}, L., {Cox}, T.~J., {Di Matteo}, T., {Robertson},
  B., \& {Springel}, V. 2006, \apjs, 163, 1

\bibitem[{{Ivezi{\'c}} {et~al.}(2004){Ivezi{\'c}}, {Lupton}, {Schlegel},
  {Boroski}, {Adelman-McCarthy}, {Yanny}, {Kent}, {Stoughton}, {Finkbeiner},
  {Padmanabhan}, {Rockosi}, {Gunn}, {Knapp}, {Strauss}, {Richards},
  {Eisenstein}, {Nicinski}, {Kleinman}, {Krzesinski}, {Newman}, {Snedden},
  {Thakar}, {Szalay}, {Munn}, {Smith}, {Tucker}, \& {Lee}}]{ivezic04}
{Ivezi{\'c}}, {\v Z}., {Lupton}, R.~H., {Schlegel}, D., {Boroski}, B.,
  {Adelman-McCarthy}, J., {Yanny}, B., {Kent}, S., {Stoughton}, C.,
  {Finkbeiner}, D., {Padmanabhan}, N., {Rockosi}, C.~M., {Gunn}, J.~E.,
  {Knapp}, G.~R., {Strauss}, M.~A., {Richards}, G.~T., {Eisenstein}, D.,
  {Nicinski}, T., {Kleinman}, S.~J., {Krzesinski}, J., {Newman}, P.~R.,
  {Snedden}, S., {Thakar}, A.~R., {Szalay}, A., {Munn}, J.~A., {Smith}, J.~A.,
  {Tucker}, D., \& {Lee}, B.~C. 2004, Astronomische Nachrichten, 325, 583

\bibitem[{{Kang} {et~al.}(2007){Kang}, {van den Bosch}, \& {Pasquali}}]{kang07}
{Kang}, X., {van den Bosch}, F.~C., \& {Pasquali}, A. 2007, \mnras, 808

\bibitem[{{Kauffmann} {et~al.}(2003){Kauffmann}, {Heckman}, {White}, {Charlot},
  {Tremonti}, {Peng}, {Seibert}, {Brinkmann}, {Nichol}, {SubbaRao}, \&
  {York}}]{kauffmann03b}
{Kauffmann}, G., {Heckman}, T.~M., {White}, S.~D.~M., {Charlot}, S.,
  {Tremonti}, C., {Peng}, E.~W., {Seibert}, M., {Brinkmann}, J., {Nichol},
  R.~C., {SubbaRao}, M., \& {York}, D. 2003, \mnras, 341, 54

\bibitem[{{Kaviraj} \& et~al.(2007)}]{kaviraj07}
{Kaviraj}, S., \& et~al. 2007, submitted, (astro-ph/0709.0806)

\bibitem[{{Khochfar} \& {Burkert}(2003)}]{khochfar03}
{Khochfar}, S., \& {Burkert}, A. 2003, \apjl, 597, L117

\bibitem[{{Khochfar} \& {Burkert}(2005)}]{khochfar05}
---. 2005, \mnras, 359, 1379

\bibitem[{{Kormendy} \& {Bender}(1996)}]{kormendy96}
{Kormendy}, J., \& {Bender}, R. 1996, \apjl, 464, L119+

\bibitem[{{Kroupa}(2001)}]{kroupa01}
{Kroupa}, P. 2001, \mnras, 322, 231

\bibitem[{{Lauer}(1986)}]{lauer86b}
{Lauer}, T.~R. 1986, \apj, 311, 34

\bibitem[{{Lauer}(1988)}]{lauer88}
---. 1988, \apj, 325, 49

\bibitem[{{Lauer} {et~al.}(2007){Lauer}, {Faber}, {Richstone}, {Gebhardt},
  {Tremaine}, {Postman}, {Dressler}, {Aller}, {Filippenko}, {Green}, {Ho},
  {Kormendy}, {Magorrian}, \& {Pinkney}}]{lauer07a}
{Lauer}, T.~R., {Faber}, S.~M., {Richstone}, D., {Gebhardt}, K., {Tremaine},
  S., {Postman}, M., {Dressler}, A., {Aller}, M.~C., {Filippenko}, A.~V.,
  {Green}, R., {Ho}, L.~C., {Kormendy}, J., {Magorrian}, J., \& {Pinkney}, J.
  2007, \apj, 662, 808

\bibitem[{{Le F{\` e}vre} {et~al.}(2000){Le F{\` e}vre}, {Abraham}, {Lilly},
  {Ellis}, {Brinchmann}, {Schade}, {Tresse}, {Colless}, {Crampton},
  {Glazebrook}, {Hammer}, \& {Broadhurst}}]{lefevre00}
{Le F{\` e}vre}, O., {Abraham}, R., {Lilly}, S.~J., {Ellis}, R.~S.,
  {Brinchmann}, J., {Schade}, D., {Tresse}, L., {Colless}, M., {Crampton}, D.,
  {Glazebrook}, K., {Hammer}, F., \& {Broadhurst}, T. 2000, \mnras, 311, 565

\bibitem[{{Lin} {et~al.}(2004){Lin}, {Koo}, {Willmer}, {Patton}, {Conselice},
  {Yan}, {Coil}, {Cooper}, {Davis}, {Faber}, {Gerke}, {Guhathakurta}, \&
  {Newman}}]{lin04}
{Lin}, L., {Koo}, D.~C., {Willmer}, C.~N.~A., {Patton}, D.~R., {Conselice},
  C.~J., {Yan}, R., {Coil}, A.~L., {Cooper}, M.~C., {Davis}, M., {Faber},
  S.~M., {Gerke}, B.~F., {Guhathakurta}, P., \& {Newman}, J.~A. 2004, \apjl,
  617, L9

\bibitem[{{Lotz} \& et~al.(2006)}]{lotz06}
{Lotz}, J., \& et~al. 2006, \apj \, submitted, (astro-ph/0602088)

\bibitem[{{Lupton} {et~al.}(2002){Lupton}, {Ivezic}, {Gunn}, {Knapp},
  {Strauss}, \& {Yasuda}}]{lupton02}
{Lupton}, R.~H., {Ivezic}, Z., {Gunn}, J.~E., {Knapp}, G., {Strauss}, M.~A., \&
  {Yasuda}, N. 2002, 4836, 350

\bibitem[{{Malin} \& {Carter}(1983)}]{malin83}
{Malin}, D.~F., \& {Carter}, D. 1983, \apj, 274, 534

\bibitem[{{Maller} {et~al.}(2006){Maller}, {Katz}, {Kere{\v s}}, {Dav{\'e}}, \&
  {Weinberg}}]{maller06}
{Maller}, A.~H., {Katz}, N., {Kere{\v s}}, D., {Dav{\'e}}, R., \& {Weinberg},
  D.~H. 2006, \apj, 647, 763

\bibitem[{{Masjedi} {et~al.}(2007){Masjedi}, {Hogg}, \& {Blanton}}]{masjedi07}
{Masjedi}, M., {Hogg}, D.~W., \& {Blanton}, M. 2007, \apj \, submitted,
  (astro-ph/0708.3240)

\bibitem[{{Masjedi} {et~al.}(2006){Masjedi}, {Hogg}, {Cool}, {Eisenstein},
  {Blanton}, {Zehavi}, {Berlind}, {Bell}, {Schneider}, {Warren}, \&
  {Brinkmann}}]{masjedi06}
{Masjedi}, M., {Hogg}, D.~W., {Cool}, R.~J., {Eisenstein}, D.~J., {Blanton},
  M.~R., {Zehavi}, I., {Berlind}, A.~A., {Bell}, E.~F., {Schneider}, D.~P.,
  {Warren}, M.~S., \& {Brinkmann}, J. 2006, \apj, 644, 54

\bibitem[{{McIntosh} {et~al.}(2005){McIntosh}, {Bell}, {Rix}, {Wolf},
  {Heymans}, {Peng}, {Somerville}, {Barden}, {Beckwith}, {Borch}, {Caldwell},
  {H{\"a}u{\ss}ler}, {Jahnke}, {Jogee}, {Meisenheimer}, {S{\'a}nchez}, \&
  {Wisotzki}}]{mcintosh05b}
{McIntosh}, D.~H., {Bell}, E.~F., {Rix}, H.-W., {Wolf}, C., {Heymans}, C.,
  {Peng}, C.~Y., {Somerville}, R.~S., {Barden}, M., {Beckwith}, S.~V.~W.,
  {Borch}, A., {Caldwell}, J.~A.~R., {H{\"a}u{\ss}ler}, B., {Jahnke}, K.,
  {Jogee}, S., {Meisenheimer}, K., {S{\'a}nchez}, S.~F., \& {Wisotzki}, L.
  2005, \apj, 632, 191

\bibitem[{{Merritt}(1985)}]{merritt85}
{Merritt}, D. 1985, \apj, 289, 18

\bibitem[{{Mihos}(2001)}]{mihos01}
{Mihos}, J.~C. 2001, \apj, 550, 94

\bibitem[{{Mihos} {et~al.}(2005){Mihos}, {Harding}, {Feldmeier}, \&
  {Morrison}}]{mihos05}
{Mihos}, J.~C., {Harding}, P., {Feldmeier}, J., \& {Morrison}, H. 2005, \apjl,
  631, L41

\bibitem[{{Mo} \& {White}(2002)}]{mo02}
{Mo}, H.~J., \& {White}, S.~D.~M. 2002, \mnras, 336, 112

\bibitem[{{Monaco} {et~al.}(2006){Monaco}, {Murante}, {Borgani}, \&
  {Fontanot}}]{monaco06}
{Monaco}, P., {Murante}, G., {Borgani}, S., \& {Fontanot}, F. 2006, \apjl, 652,
  L89

\bibitem[{{Naab} \& {Burkert}(2003)}]{naab03}
{Naab}, T., \& {Burkert}, A. 2003, \apj, 597, 893

\bibitem[{{Naab} {et~al.}(1999){Naab}, {Burkert}, \& {Hernquist}}]{naab99}
{Naab}, T., {Burkert}, A., \& {Hernquist}, L. 1999, \apjl, 523, L133

\bibitem[{{Naab} {et~al.}(2006{\natexlab{a}}){Naab}, {Jesseit}, \&
  {Burkert}}]{naab06d}
{Naab}, T., {Jesseit}, R., \& {Burkert}, A. 2006{\natexlab{a}}, \mnras, 372,
  839

\bibitem[{{Naab} {et~al.}(2006{\natexlab{b}}){Naab}, {Khochfar}, \&
  {Burkert}}]{naab06a}
{Naab}, T., {Khochfar}, S., \& {Burkert}, A. 2006{\natexlab{b}}, \apjl, 636,
  L81

\bibitem[{{Ostriker} \& {Tremaine}(1975)}]{ostriker75}
{Ostriker}, J.~P., \& {Tremaine}, S.~D. 1975, \apjl, 202, L113

\bibitem[{{Patton} {et~al.}(2000){Patton}, {Carlberg}, {Marzke}, {Pritchet},
  {da Costa}, \& {Pellegrini}}]{patton00}
{Patton}, D.~R., {Carlberg}, R.~G., {Marzke}, R.~O., {Pritchet}, C.~J., {da
  Costa}, L.~N., \& {Pellegrini}, P.~S. 2000, \apj, 536, 153

\bibitem[{{Patton} {et~al.}(2002){Patton}, {Pritchet}, {Carlberg}, {Marzke},
  {Yee}, {Hall}, {Lin}, {Morris}, {Sawicki}, {Shepherd}, \& {Wirth}}]{patton02}
{Patton}, D.~R., {Pritchet}, C.~J., {Carlberg}, R.~G., {Marzke}, R.~O., {Yee},
  H.~K.~C., {Hall}, P.~B., {Lin}, H., {Morris}, S.~L., {Sawicki}, M.,
  {Shepherd}, C.~W., \& {Wirth}, G.~D. 2002, \apj, 565, 208

\bibitem[{{Peng} {et~al.}(2002){Peng}, {Ho}, {Impey}, \& {Rix}}]{peng02}
{Peng}, C.~Y., {Ho}, L.~C., {Impey}, C.~D., \& {Rix}, H.-W. 2002, \aj, 124, 266

\bibitem[{{Pier} {et~al.}(2003){Pier}, {Munn}, {Hindsley}, {Hennessy}, {Kent},
  {Lupton}, \& {Ivezi{\' c}}}]{pier03}
{Pier}, J.~R., {Munn}, J.~A., {Hindsley}, R.~B., {Hennessy}, G.~S., {Kent},
  S.~M., {Lupton}, R.~H., \& {Ivezi{\' c}}, {\v Z}. 2003, \aj, 125, 1559

\bibitem[{{Postman} \& {Geller}(1984)}]{postman84}
{Postman}, M., \& {Geller}, M.~J. 1984, \apj, 281, 95

\bibitem[{{Rines} {et~al.}(2007){Rines}, {Finn}, \& {Vikhlinin}}]{rines07}
{Rines}, K., {Finn}, R., \& {Vikhlinin}, A. 2007, \apjl, 665, L9

\bibitem[{{Scarlata} {et~al.}(2007){Scarlata}, {Carollo}, {Lilly}, {Feldmann},
  {Kampczyk}, {Renzini}, {Cimatti}, {Halliday}, {Daddi}, {Sargent},
  {Koekemoer}, {Scoville}, {Kneib}, {Leauthaud}, {Massey}, {Rhodes}, {Tasca},
  {Capak}, {McCracken}, {Mobasher}, {Taniguchi}, {Thompson}, {Ajiki}, {Aussel},
  {Murayama}, {Sanders}, {Sasaki}, {Shioya}, \& {Takahashi}}]{scarlata07}
{Scarlata}, C., {Carollo}, C.~M., {Lilly}, S.~J., {Feldmann}, R., {Kampczyk},
  P., {Renzini}, A., {Cimatti}, A., {Halliday}, C., {Daddi}, E., {Sargent},
  M.~T., {Koekemoer}, A., {Scoville}, N., {Kneib}, J.-P., {Leauthaud}, A.,
  {Massey}, R., {Rhodes}, J., {Tasca}, L., {Capak}, P., {McCracken}, H.~J.,
  {Mobasher}, B., {Taniguchi}, Y., {Thompson}, D., {Ajiki}, M., {Aussel}, H.,
  {Murayama}, T., {Sanders}, D.~B., {Sasaki}, S., {Shioya}, Y., \& {Takahashi},
  M. 2007, \apjs, 172, 494

\bibitem[{{Schlegel} {et~al.}(1998){Schlegel}, {Finkbeiner}, \&
  {Davis}}]{schlegel98}
{Schlegel}, D.~J., {Finkbeiner}, D.~P., \& {Davis}, M. 1998, \apj, 500, 525

\bibitem[{{Schweizer} \& {Seitzer}(1992)}]{schweizer92}
{Schweizer}, F., \& {Seitzer}, P. 1992, \aj, 104, 1039

\bibitem[{{Smith} {et~al.}(2005){Smith}, {Treu}, {Ellis}, {Moran}, \&
  {Dressler}}]{smith05}
{Smith}, G.~P., {Treu}, T., {Ellis}, R.~S., {Moran}, S.~M., \& {Dressler}, A.
  2005, \apj, 620, 78

\bibitem[{{Smith} {et~al.}(2002){Smith}, {Tucker}, {Kent}, {Richmond},
  {Fukugita}, {Ichikawa}, {Ichikawa}, {Jorgensen}, {Uomoto}, {Gunn}, {Hamabe},
  {Watanabe}, {Tolea}, {Henden}, {Annis}, {Pier}, {McKay}, {Brinkmann}, {Chen},
  {Holtzman}, {Shimasaku}, \& {York}}]{smith02}
{Smith}, J.~A., {Tucker}, D.~L., {Kent}, S., {Richmond}, M.~W., {Fukugita}, M.,
  {Ichikawa}, T., {Ichikawa}, S.-i., {Jorgensen}, A.~M., {Uomoto}, A., {Gunn},
  J.~E., {Hamabe}, M., {Watanabe}, M., {Tolea}, A., {Henden}, A., {Annis}, J.,
  {Pier}, J.~R., {McKay}, T.~A., {Brinkmann}, J., {Chen}, B., {Holtzman}, J.,
  {Shimasaku}, K., \& {York}, D.~G. 2002, \aj, 123, 2121

\bibitem[{{Strateva} {et~al.}(2001){Strateva}, {Ivezi{\' c}}, {Knapp},
  {Narayanan}, {Strauss}, {Gunn}, {Lupton}, {Schlegel}, {Bahcall}, {Brinkmann},
  {Brunner}, {Budav{\' a}ri}, {Csabai}, {Castander}, {Doi}, {Fukugita}, {Gy{\H
  o}ry}, {Hamabe}, {Hennessy}, {Ichikawa}, {Kunszt}, {Lamb}, {McKay},
  {Okamura}, {Racusin}, {Sekiguchi}, {Schneider}, {Shimasaku}, \&
  {York}}]{strateva01}
{Strateva}, I., {Ivezi{\' c}}, {\v Z}., {Knapp}, G.~R., {Narayanan}, V.~K.,
  {Strauss}, M.~A., {Gunn}, J.~E., {Lupton}, R.~H., {Schlegel}, D., {Bahcall},
  N.~A., {Brinkmann}, J., {Brunner}, R.~J., {Budav{\' a}ri}, T., {Csabai}, I.,
  {Castander}, F.~J., {Doi}, M., {Fukugita}, M., {Gy{\H o}ry}, Z., {Hamabe},
  M., {Hennessy}, G., {Ichikawa}, T., {Kunszt}, P.~Z., {Lamb}, D.~Q., {McKay},
  T.~A., {Okamura}, S., {Racusin}, J., {Sekiguchi}, M., {Schneider}, D.~P.,
  {Shimasaku}, K., \& {York}, D. 2001, \aj, 122, 1861

\bibitem[{{Strauss} {et~al.}(2002){Strauss}, {Weinberg}, {Lupton}, {Narayanan},
  {Annis}, {Bernardi}, {Blanton}, {Burles}, {Connolly}, {Dalcanton}, {Doi},
  {Eisenstein}, {Frieman}, {Fukugita}, {Gunn}, {Ivezi{\' c}}, {Kent}, {Kim},
  {Knapp}, {Kron}, {Munn}, {Newberg}, {Nichol}, {Okamura}, {Quinn}, {Richmond},
  {Schlegel}, {Shimasaku}, {SubbaRao}, {Szalay}, {Vanden Berk}, {Vogeley},
  {Yanny}, {Yasuda}, {York}, \& {Zehavi}}]{strauss02}
{Strauss}, M.~A., {Weinberg}, D.~H., {Lupton}, R.~H., {Narayanan}, V.~K.,
  {Annis}, J., {Bernardi}, M., {Blanton}, M., {Burles}, S., {Connolly}, A.~J.,
  {Dalcanton}, J., {Doi}, M., {Eisenstein}, D., {Frieman}, J.~A., {Fukugita},
  M., {Gunn}, J.~E., {Ivezi{\' c}}, {\v Z}., {Kent}, S., {Kim}, R.~S.~J.,
  {Knapp}, G.~R., {Kron}, R.~G., {Munn}, J.~A., {Newberg}, H.~J., {Nichol},
  R.~C., {Okamura}, S., {Quinn}, T.~R., {Richmond}, M.~W., {Schlegel}, D.~J.,
  {Shimasaku}, K., {SubbaRao}, M., {Szalay}, A.~S., {Vanden Berk}, D.,
  {Vogeley}, M.~S., {Yanny}, B., {Yasuda}, N., {York}, D.~G., \& {Zehavi}, I.
  2002, \aj, 124, 1810

\bibitem[{{Toomre}(1977)}]{toomre77}
{Toomre}, A. 1977, in Evolution of Galaxies and Stellar Populations, 401--+

\bibitem[{{Toomre} \& {Toomre}(1972)}]{toomre72}
{Toomre}, A., \& {Toomre}, J. 1972, \apj, 178, 623

\bibitem[{{Tran} {et~al.}(2005){Tran}, {van Dokkum}, {Franx}, {Illingworth},
  {Kelson}, \& {Schreiber}}]{tran05}
{Tran}, K.-V.~H., {van Dokkum}, P., {Franx}, M., {Illingworth}, G.~D.,
  {Kelson}, D.~D., \& {Schreiber}, N.~M.~F. 2005, \apjl, 627, L25

\bibitem[{{Tucker} {et~al.}(2006){Tucker}, {Kent}, {Richmond}, {Annis},
  {Smith}, {Allam}, {Rodgers}, {Stute}, {Adelman-McCarthy}, {Brinkmann}, {Doi},
  {Finkbeiner}, {Fukugita}, {Goldston}, {Greenway}, {Gunn}, {Hendry}, {Hogg},
  {Ichikawa}, {Ivezi{\'c}}, {Knapp}, {Lampeitl}, {Lee}, {Lin}, {McKay},
  {Merrelli}, {Munn}, {Neilsen}, {Newberg}, {Richards}, {Schlegel},
  {Stoughton}, {Uomoto}, \& {Yanny}}]{tucker06}
{Tucker}, D.~L., {Kent}, S., {Richmond}, M.~W., {Annis}, J., {Smith}, J.~A.,
  {Allam}, S.~S., {Rodgers}, C.~T., {Stute}, J.~L., {Adelman-McCarthy}, J.~K.,
  {Brinkmann}, J., {Doi}, M., {Finkbeiner}, D., {Fukugita}, M., {Goldston}, J.,
  {Greenway}, B., {Gunn}, J.~E., {Hendry}, J.~S., {Hogg}, D.~W., {Ichikawa},
  S.-I., {Ivezi{\'c}}, {\v Z}., {Knapp}, G.~R., {Lampeitl}, H., {Lee}, B.~C.,
  {Lin}, H., {McKay}, T.~A., {Merrelli}, A., {Munn}, J.~A., {Neilsen}, Jr.,
  E.~H., {Newberg}, H.~J., {Richards}, G.~T., {Schlegel}, D.~J., {Stoughton},
  C., {Uomoto}, A., \& {Yanny}, B. 2006, Astronomische Nachrichten, 327, 821

\bibitem[{{van Dokkum}(2005)}]{vandokkum05}
{van Dokkum}, P.~G. 2005, \aj, 130, 2647

\bibitem[{{van Dokkum} {et~al.}(1999){van Dokkum}, {Franx}, {Fabricant},
  {Kelson}, \& {Illingworth}}]{vandokkum99}
{van Dokkum}, P.~G., {Franx}, M., {Fabricant}, D., {Kelson}, D.~D., \&
  {Illingworth}, G.~D. 1999, \apjl, 520, L95

\bibitem[{{Wake} {et~al.}(2006){Wake}, {Nichol}, {Eisenstein}, {Loveday},
  {Edge}, {Cannon}, {Smail}, {Schneider}, {Scranton}, {Carson}, {Ross},
  {Brunner}, {Colless}, {Couch}, {Croom}, {Driver}, {da {\^A}ngela}, {Jester},
  {de Propris}, {Drinkwater}, {Bland-Hawthorn}, {Pimbblet}, {Roseboom},
  {Shanks}, {Sharp}, \& {Brinkmann}}]{wake06}
{Wake}, D.~A., {Nichol}, R.~C., {Eisenstein}, D.~J., {Loveday}, J., {Edge},
  A.~C., {Cannon}, R., {Smail}, I., {Schneider}, D.~P., {Scranton}, R.,
  {Carson}, D., {Ross}, N.~P., {Brunner}, R.~J., {Colless}, M., {Couch}, W.~J.,
  {Croom}, S.~M., {Driver}, S.~P., {da {\^A}ngela}, J., {Jester}, S., {de
  Propris}, R., {Drinkwater}, M.~J., {Bland-Hawthorn}, J., {Pimbblet}, K.~A.,
  {Roseboom}, I.~G., {Shanks}, T., {Sharp}, R.~G., \& {Brinkmann}, J. 2006,
  \mnras, 372, 537

\bibitem[{{Weinberg}(1986)}]{weinberg86}
{Weinberg}, M.~D. 1986, \apj, 300, 93

\bibitem[{{Weinmann} {et~al.}(2006){Weinmann}, {van den Bosch}, {Yang}, \&
  {Mo}}]{weinmann06a}
{Weinmann}, S.~M., {van den Bosch}, F.~C., {Yang}, X., \& {Mo}, H.~J. 2006,
  \mnras, 366, 2

\bibitem[{{White} {et~al.}(2007){White}, {Zheng}, {Brown}, {Dey}, \&
  {Jannuzi}}]{white07}
{White}, M., {Zheng}, Z., {Brown}, M.~J.~I., {Dey}, A., \& {Jannuzi}, B.~T.
  2007, \apjl, 655, L69

\bibitem[{{Yang} {et~al.}(2005){Yang}, {Mo}, {van den Bosch}, \&
  {Jing}}]{yang05a}
{Yang}, X., {Mo}, H.~J., {van den Bosch}, F.~C., \& {Jing}, Y.~P. 2005, \mnras,
  356, 1293

\bibitem[{{York} {et~al.}(2000){York}, {Adelman}, {Anderson}, {Anderson},
  {Annis}, {Bahcall}, {Bakken}, {Barkhouser}, {Bastian}, {Berman}, \& {et
  al.}}]{york00}
{York}, D.~G., {Adelman}, J., {Anderson}, J.~E., {Anderson}, S.~F., {Annis},
  J., {Bahcall}, N.~A., {Bakken}, J.~A., {Barkhouser}, R., {Bastian}, S.,
  {Berman}, E., \& {et al.} 2000, \aj, 120, 1579

\bibitem[{{Zehavi} {et~al.}(2002){Zehavi}, {Blanton}, {Frieman}, {Weinberg},
  {Mo}, {Strauss}, {Anderson}, {Annis}, {Bahcall}, {Bernardi}, {Briggs},
  {Brinkmann}, {Burles}, {Carey}, {Castander}, {Connolly}, {Csabai},
  {Dalcanton}, {Dodelson}, {Doi}, {Eisenstein}, {Evans}, {Finkbeiner},
  {Friedman}, {Fukugita}, {Gunn}, {Hennessy}, {Hindsley}, {Ivezi{\'c}}, {Kent},
  {Knapp}, {Kron}, {Kunszt}, {Lamb}, {Leger}, {Long}, {Loveday}, {Lupton},
  {McKay}, {Meiksin}, {Merrelli}, {Munn}, {Narayanan}, {Newcomb}, {Nichol},
  {Owen}, {Peoples}, {Pope}, {Rockosi}, {Schlegel}, {Schneider}, {Scoccimarro},
  {Sheth}, {Siegmund}, {Smee}, {Snir}, {Stebbins}, {Stoughton}, {SubbaRao},
  {Szalay}, {Szapudi}, {Tegmark}, {Tucker}, {Uomoto}, {Vanden Berk}, {Vogeley},
  {Waddell}, {Yanny}, \& {York}}]{zehavi02}
{Zehavi}, I., {Blanton}, M.~R., {Frieman}, J.~A., {Weinberg}, D.~H., {Mo},
  H.~J., {Strauss}, M.~A., {Anderson}, S.~F., {Annis}, J., {Bahcall}, N.~A.,
  {Bernardi}, M., {Briggs}, J.~W., {Brinkmann}, J., {Burles}, S., {Carey}, L.,
  {Castander}, F.~J., {Connolly}, A.~J., {Csabai}, I., {Dalcanton}, J.~J.,
  {Dodelson}, S., {Doi}, M., {Eisenstein}, D., {Evans}, M.~L., {Finkbeiner},
  D.~P., {Friedman}, S., {Fukugita}, M., {Gunn}, J.~E., {Hennessy}, G.~S.,
  {Hindsley}, R.~B., {Ivezi{\'c}}, {\v Z}., {Kent}, S., {Knapp}, G.~R., {Kron},
  R., {Kunszt}, P., {Lamb}, D.~Q., {Leger}, R.~F., {Long}, D.~C., {Loveday},
  J., {Lupton}, R.~H., {McKay}, T., {Meiksin}, A., {Merrelli}, A., {Munn},
  J.~A., {Narayanan}, V., {Newcomb}, M., {Nichol}, R.~C., {Owen}, R.,
  {Peoples}, J., {Pope}, A., {Rockosi}, C.~M., {Schlegel}, D., {Schneider},
  D.~P., {Scoccimarro}, R., {Sheth}, R.~K., {Siegmund}, W., {Smee}, S., {Snir},
  Y., {Stebbins}, A., {Stoughton}, C., {SubbaRao}, M., {Szalay}, A.~S.,
  {Szapudi}, I., {Tegmark}, M., {Tucker}, D.~L., {Uomoto}, A., {Vanden Berk},
  D., {Vogeley}, M.~S., {Waddell}, P., {Yanny}, B., \& {York}, D.~G. 2002,
  \apj, 571, 172

\bibitem[{{Zibetti} {et~al.}(2005){Zibetti}, {White}, {Schneider}, \&
  {Brinkmann}}]{zibetti05}
{Zibetti}, S., {White}, S.~D.~M., {Schneider}, D.~P., \& {Brinkmann}, J. 2005,
  \mnras, 358, 949

\end{thebibliography}


\onecolumn

\begin{deluxetable}{lcccc}
\tablewidth{0pt}
\tablenum{1}
\tabletypesize{\small}
\tablecolumns{5}
\tablecaption{Massive Galaxy Content in sampM and the $z\leq0.12$ SDSS DR2 Volume}
\tablehead{\colhead{} & \multicolumn{4}{c}{stellar-mass bins} \\
\cline{2-5} \\ 
\colhead{} & \colhead{[10.7,11.0]} & \colhead{[11.0,11.3]} & \colhead{[11.3,11.6]} & \colhead{[11.6,11.9]}}
\startdata
Total in DR2 volume & 28377 & 10690 & 1943 & 165 \\
Red sequence in DR2 volume & 23657 & 9846 & 1897 & 164 \\
Red percent & 83.4\% & 92.1\% & 97.6\% & 99.4\% \\
\cline{1-5} \\ 
Total in sampM & 3238 & 1415 & 599 & 120 \\
Red sequence in sampM & 2979 & 1329 & 586 & 119 \\
Red percent & 92.0\% & 93.9\% & 97.8\% & 99.2\% \\
\cline{1-5} \\ 
Centrals in sampM & 29 & 241 & 460 & 115 \\
Percent of DR2 total & 0.1\% & 2.3\% & 23.7\% & 69.7\% \\
\cline{1-5} \\ 
Satellites in sampM & 3209 & 1174 & 139 & 5 \\
Percent of DR2 total & 11.3\% & 11.0\% & 7.2\% & 3.0\% \\
\enddata
\tablecomments{Bins of stellar mass are in units of 
$\log_{10}({\rm M}_{\sun})$.}
\label{mstar_counts}
\end{deluxetable}

\begin{deluxetable}{rccrrrccc}
\tablewidth{0pt}
\tablenum{2}
\tabletypesize{\small}
\tablecolumns{9}
\tablecaption{Progenitors of Massive Merger Systems}
\tablehead{\colhead{Group ID} & \colhead{${\rm M}_{\rm halo}$} & \colhead{Flag} & \colhead{NYU ID} & \colhead{R.A.} & \colhead{Dec} & \colhead{$z$} & \colhead{${\rm M}_{\rm star}$} & \colhead{$^{0.0}(g-r)$} \\
\colhead{(1)} & \colhead{(2)} & \colhead{(3)} & \colhead{(4)} & \colhead{(5)} & \colhead{(6)} & \colhead{(7)} & \colhead{(8)} & \colhead{(9)}}
\startdata
   121 & $13.74$ & CEN & 150206 &  14:14:32.6 & +01:43:53.6 & $0.053$ & $11.21(11.28)$ & $0.82(0.84)$ \\
       &         & no &     na &  14:14:32.6 & +01:44:01.5 & na & $10.98(10.95)$ & $0.78(0.77)$ \\
   830 & $13.55$ & CEN & 249473 &  08:54:58.9 & +49:08:32.4 & $0.052$ & $11.31(11.47)$ & $0.86(0.93)$ \\
 & & no &     na &  08:55:00.6 & +49:08:32.8 & na & $10.83(10.82)$ & $0.69(0.74)$ \\
   419 & $14.41$ & CEN &   9993 &  12:27:37.1 & -00:23:02.4 & $0.115$ & $11.54(11.72)$ & $0.78(0.85)$ \\
 & & no &     na &  12:27:36.7 & -00:23:10.8 & na & $11.11(11.17)$ & $0.79(0.79)$ \\
    54 & $14.62$ & CEN &  11349 &  15:08:25.8 & -00:15:58.6 & $0.090$ & $11.63(11.65)$ & $0.81(0.80)$ \\
 & & no &     na &  15:08:25.0 & -00:16:07.1 & na & $11.49(11.45)$ & $0.81(0.80)$ \\
   614 & $13.82$ & CEN & 124158 &  13:52:02.2 & +66:50:20.1 & $0.068$ & $11.31(11.37)$ & $0.82(0.85)$ \\
 & & no &     na &  13:52:01.0 & +66:50:19.3 & na & $10.93(11.02)$ & $0.70(0.75)$ \\
   163 & $14.47$ & CEN & 175344 &  15:09:59.4 & +03:00:11.1 & $0.092$ & $11.51(11.58)$ & $0.81(0.84)$ \\
 & & no &     na &  15:09:59.6 & +03:00:03.8 & na & $11.52(11.21)^{\ddagger}$ & $1.05(0.87)$ \\
   539 & $14.30$ & CEN & 222852 &  00:56:20.1 & -09:36:29.7 & $0.103$ & $11.51(11.51)$ & $0.77(0.77)$ \\
 & & no &     na &  00:56:20.0 & -09:36:33.7 & na & $11.25(11.34)$ & $0.74(0.72)$ \\
   393 & $14.26$ & CEN & 261132 &  10:04:39.4 & +02:57:42.8 & $0.104$ & $11.37(11.41)$ & $0.80(0.79)$ \\
 & & no &     na &  10:04:39.5 & +02:57:39.9 & na & $10.73(10.73)$ & $0.66(0.62)$ \\
   398 & $14.29$ & CEN & 293645 &  10:37:29.8 & -00:40:40.5 & $0.096$ & $11.33(11.37)$ & $0.75(0.79)$ \\
 & & no &     na &  10:37:29.9 & -00:40:46.3 & na & $11.23(11.11)$ & $0.88(0.84)$ \\
   214 & $14.18$ & CEN & 311008 &  15:41:35.7 & +55:58:39.8 & $0.068$ & $11.31(11.30)$ & $0.89(0.87)$ \\
 & & no &     na &  15:41:34.5 & +55:58:38.9 & na & $11.26(11.24)$ & $0.76(0.76)$ \\
   291 & $14.07$ & CEN & 392792 &  22:28:25.5 & -09:37:22.3 & $0.083$ & $11.35(11.42)$ & $0.76(0.81)$ \\
 & & no &     na &  22:28:25.6 & -09:37:30.4 & na & $11.28(11.42)$ & $0.81(0.86)$ \\
     5 & $14.24$ & CEN & 301558 &  14:40:42.8 & +03:27:55.5 & $0.027$ & $11.39(11.48)$ & $0.82(0.81)$ \\
 & & SAT & 301560 &  14:40:39.0 & +03:28:11.0 & $0.027$ & $11.14(11.49)^{\ddagger}$ & $0.87(0.82)$ \\
   102 & $14.32$ & CEN &  44192 &  09:58:52.2 & +01:03:33.1 & $0.081$ & $11.26(11.33)$ & $0.67(0.74)$ \\
 & & SAT &  44193 &  09:58:52.0 & +01:03:46.4 & $0.082$ & $11.45(11.68)$ & $1.21(1.06)$ \\
   759 & $14.13$ & CEN &  88664 &  08:46:13.1 & +53:26:38.1 & $0.113$ & $11.73(11.43)^{\ddagger}$ & $1.13(0.92)$ \\
 & & SAT &  88665 &  08:46:13.3 & +53:26:35.9 & $0.113$ & $10.84(11.17)^{\ddagger}$ & $0.66(0.69)$ \\
    74 & $14.27$ & CEN & 258681 &  11:45:37.2 & +64:30:41.4 & $0.063$ & $11.42(11.46)$ & $0.84(0.85)$ \\
 & & SAT & 258682 &  11:45:37.4 & +64:30:45.3 & $0.064$ & $11.37(11.41)$ & $0.90(0.81)$ \\
   847 & $13.91$ & CEN & 274752 &  10:34:09.7 & +04:21:29.8 & $0.100$ & $11.42(11.52)$ & $0.82(0.84)$ \\
 & & SAT & 274751 &  10:34:09.1 & +04:21:30.8 & $0.100$ & $11.25(11.56)^{\ddagger}$ & $0.86(0.88)$ \\
   572 & $13.92$ & CEN & 371303 &  13:30:10.3 & -02:06:18.0 & $0.087$ & $11.35(11.38)$ & $0.81(0.79)$ \\
 & & SAT & 371304 &  13:30:10.9 & -02:06:13.6 & $0.086$ & $11.38(11.43)$ & $0.87(0.84)$ \\
  1775 & $13.98$ & CEN &  92509 &  17:20:36.1 & +56:39:42.5 & $0.120$ & $11.40(11.42)$ & $0.82(0.83)$ \\
 & & SAT &  92510 &  17:20:37.7 & +56:39:45.1 & $0.120$ & $11.33(11.39)$ & $0.78(0.80)$ \\
  1545 & $13.43$ & SAT & 364190 &  13:36:43.6 & -03:29:57.0 & $0.053$ & $10.91(10.79)$ & $1.01(0.93)$ \\
 & & no$^{\dagger}$ &     na &  13:36:44.3 & -03:29:52.5 & na & $11.05(11.15)$ & $0.72(0.79)$ \\
   126 & $13.55$ & SAT & 367419 &  13:59:25.2 & -03:12:29.0 & $0.025$ & $10.73(10.58)$ & $0.95(0.86)$ \\
 & & no &     na &  13:59:24.8 & -03:12:33.1 & na & $11.07(11.13)$ & $0.81(0.83)$ \\
    37 & $14.72$ & SAT$^{\dagger\dagger}$ &  33684 &  15:11:20.3 & -00:07:20.1 & $0.089$ & $11.02(11.10)$ & $0.74(0.77)$ \\
 & & no$^{\dagger\dagger}$ &     na &  15:11:19.2 & -00:07:16.5 & na & $11.07(11.03)$ & $0.79(0.74)$ \\
    72 & $14.79$ & SAT &  83539 &  08:54:48.7 & +00:51:02.6 & $0.107$ & $11.11(11.15)$ & $0.62(0.67)$ \\
 & & no &     na &  08:54:48.2 & +00:50:46.6 & na & $10.82(11.19)^{\ddagger}$ & $0.41(0.65)$ \\
  2955 & $13.81$ & SAT & 206506 &  20:45:09.4 & -06:17:05.5 & $0.112$ & $10.88(10.87)$ & $0.79(0.80)$ \\
 & & no &     na &  20:45:08.9 & -06:17:01.5 & na & $10.85(10.87)$ & $0.88(0.86)$ \\
    81 & $14.37$ & SAT & 218908 &  23:37:05.4 & +15:55:58.5 & $0.066$ & $10.94(11.14)$ & $0.68(0.74)$ \\
 & & no &     na &  23:37:06.2 & +15:56:03.2 & na & $10.22(10.43)$ & $0.49(0.57)$ \\
   219 & $14.20$ & SAT$^{\dagger\dagger}$ & 223211 &  23:54:59.6 & -09:14:49.4 & $0.074$ & $11.03(11.00)$ & $0.84(0.81)$ \\
 & & no$^{\dagger\dagger}$ &     na &  23:54:59.7 & -09:14:53.0 & na & $10.99(11.03)$ & $0.75(0.75)$ \\
    14 & $14.83$ & SAT & 278870 &  10:39:39.0 & +05:10:31.3 & $0.068$ & $10.73(10.86)$ & $0.52(0.64)$ \\
 & & no &     na &  10:39:38.7 & +05:10:32.6 & na & $11.30(11.36)$ & $0.84(0.81)$ \\
  1786 & $13.99$ & SAT & 284077 &  14:31:09.6 & +60:41:18.4 & $0.113$ & $10.74(10.72)$ & $0.81(0.79)$ \\
 & & no &     na &  14:31:10.2 & +60:41:35.7 & na & $11.05(11.18)$ & $0.74(0.81)$ \\
   344 & $13.92$ & SAT & 333778 &  12:40:30.2 & +05:52:21.5 & $0.075$ & $11.59(11.36)$ & $1.36(1.15)$ \\
 & & no &     na &  12:40:30.9 & +05:52:10.6 & na & $11.50(11.20)^{\ddagger}$ & $1.05(0.86)$ \\
   261 & $14.26$ & SAT & 336039 &  17:01:52.2 & +35:02:54.9 & $0.107$ & $11.01(11.06)$ & $0.67(0.74)$ \\
 & & no$^{\dagger}$ &     na &  17:01:53.1 & +35:03:04.0 & na & $11.43(11.47)$ & $0.75(0.87)$ \\
    75 & $14.85$ & SAT & 346478 &  12:47:56.7 & +62:36:27.6 & $0.107$ & $11.33(11.28)$ & $0.90(0.85)$ \\
 & & no &     na &  12:47:56.7 & +62:36:23.5 & na & $10.87(10.58)$ & $0.70(0.63)$ \\
   170 & $13.88$ & SAT$^{\dagger\dagger}$ & 352171 &  13:33:03.2 & +60:07:00.0 & $0.072$ & $11.37(11.06)^{\ddagger}$ & $1.13(0.94)$ \\
 & & no$^{\dagger\dagger}$ &     na &  13:33:03.4 & +60:07:03.7 & na & $11.08(10.94)$ & $0.78(0.73)$ \\
   479 & $14.28$ & SAT & 373137 &  14:09:59.4 & -01:32:18.9 & $0.117$ & $11.64(11.24)^{\ddagger}$ & $1.17(0.90)$ \\
 & & no$^{\dagger}$ &     na &  14:09:59.5 & -01:32:22.8 & na & $11.41(11.51)$ & $0.71(0.74)$ \\
  1047 & $13.97$ & SAT & 393494 &  22:22:48.8 & -09:02:14.4 & $0.084$ & $11.15(11.23)$ & $0.81(0.80)$ \\
 & & no &     na &  22:22:49.0 & -09:02:22.2 & na & $11.28(11.35)$ & $0.78(0.81)$ \\
   462 & $13.91$ & SAT & 250588 &  08:36:45.9 & +47:22:10.2 & $0.053$ & $11.13(11.15)$ & $0.81(0.81)$ \\
 & & SAT & 250589 &  08:36:44.8 & +47:22:18.9 & $0.053$ & $10.97(11.15)$ & $0.79(0.81)$ \\
   714 & $13.60$ & SAT & 604118 &  15:28:12.7 & +42:55:47.7 & $0.019$ & $10.90(10.94)$ & $0.82(0.87)$ \\
 & & SAT & 604117 &  15:28:16.7 & +42:56:38.8 & $0.018$ & $10.71(10.91)$ & $0.84(0.85)$ \\
   460 & $14.24$ & SAT & 241625 &  09:55:39.5 & +01:35:48.4 & $0.099$ & $11.24(11.20)$ & $0.80(0.76)$ \\
 & & SAT & 241629 &  09:55:40.2 & +01:35:50.3 & $0.099$ & $10.95(11.29)^{\ddagger}$ & $0.68(0.84)$ \\
   465 & $14.13$ & SAT$^{\dagger\dagger}$ & 294450 &  10:50:25.4 & -00:20:11.1 & $0.096$ & $11.20(11.30)$ & $0.80(0.84)$ \\
 & & SAT$^{\dagger\dagger}$ & 294451 &  10:50:25.5 & -00:20:10.1 & $0.093$ & $11.21(11.26)$ & $0.86(0.81)$ \\
   337 & $14.12$ & SAT & 269340 &  09:22:22.2 & +02:35:09.3 & $0.088$ & $11.19(11.22)$ & $0.83(0.82)$ \\
 & & SAT & 269341 &  09:22:22.0 & +02:35:13.8 & $0.087$ & $10.73(11.16)^{\ddagger}$ & $0.75(0.77)$ \\
\enddata
\tablecomments{For each merger pair the progenitor properties are listed
on two separate lines with the following columns: group ID number (1) and
dark-matter halo mass estimate in units of $\log_{10}({\rm M}_{\sun})$ (2) from
the public SDSS DR2 group catalog of Yang et al.; flag (3) for whether
galaxy was identified in group catalog as a central (CEN), satellite (SAT),
or not identified (no) owing to no spectroscopic redshift;
ID number (4), epoch J2000.0 celestial coordinates (5,6), and spectroscopic
redshift (7) from the NYU-VAGC; stellar mass estimates in units of 
$\log_{10}({\rm M}_{\sun})$ (8) based on SDSS Petrosian(Model) photometry
and \citet{bell03b} M/L ratios; rest-frame $K$-corrected to $z=0.0$ color (9)
from SDSS Petrosian(Model) photometry. \nl
$^{\dagger}$ Estimated stellar mass of the companion exceeds 
that of the spectroscopic CEN galaxy of the host;
the merger is added to the CEN-SAT subset in the analysis. \nl
$^{\dagger\dagger}$ Total estimated stellar mass of the two SATs 
(${\rm M}_1+{\rm M}_2$) exceeds that of the spectroscopic CEN galaxy of the
host; including/excluding the merger to the CEN-SAT subset is analyzed. \nl
$^{\ddagger}$ More than factor of 2 difference (0.3 dex) between Petrosian
and Model-based ${\rm M}_{\rm star}$ estimates.}
\label{tab:dp}
\end{deluxetable}

\end{document}